\documentclass[acmlarge]{acmart}

\AtBeginDocument{
  \providecommand\BibTeX{{
    \normalfont B\kern-0.5em{\scshape i\kern-0.25em b}\kern-0.8em\TeX}}}

\setcopyright{acmcopyright}
\copyrightyear{2020}
\acmYear{2020}
\acmDOI{10.1145/1122445.1122456}

\acmJournal{CSUR}
\acmVolume{1}
\acmNumber{1}
\acmArticle{1}
\acmMonth{1}

\usepackage{color}
\newcommand{\todo}[1]{}
\renewcommand{\todo}[1]{{\color{red} TODO: {#1}}}

\newcommand{\point}[1]{\vspace{5pt}\noindent\textbf{#1}}

\usepackage{graphicx}
\usepackage{subfigure}
\graphicspath{figures/}
\DeclareGraphicsExtensions{.pdf,.jpeg,.png}

\usepackage{pgf-pie} 
\usepackage{pgfplots}
\usepackage{bbding}
\usepackage{fancybox}
\usepackage{amsfonts}
\usepackage{amsmath}
\usepackage{fancyvrb}
\usepackage{color}
\usepackage{algorithmic}
\usepackage{framed}
\usepackage{array}
\usepackage{multirow}
\usepackage{booktabs}
\usepackage{balance}
\usepackage{flushend}

\usepackage{colortbl}
\usepackage{xcolor}

\usepackage{tikz}
\usetikzlibrary{positioning} 
\usepackage{xcolor} 

\usepackage{makecell}

\PassOptionsToPackage{hyphens}{url}
\usepackage{hyperref}
\hypersetup{colorlinks=true,allcolors=blue}

\newcommand{\PreserveBackslash}[1]{\let\temp=\\#1\let\\=\temp}
\newcolumntype{C}[1]{>{\PreserveBackslash\centering}p{#1}}
\newcolumntype{R}[1]{>{\PreserveBackslash\raggedleft}p{#1}}
\newcolumntype{L}[1]{>{\PreserveBackslash\raggedright}p{#1}}

\definecolor{gray50}{gray}{.5}
\definecolor{gray40}{gray}{.6}
\definecolor{gray30}{gray}{.7}
\definecolor{gray20}{gray}{.8}
\definecolor{gray10}{gray}{.9}
\definecolor{gray05}{gray}{.95}

\newlength\Linewidth
\def\findlength{\setlength\Linewidth\linewidth
\addtolength\Linewidth{-4\fboxrule}
\addtolength\Linewidth{-3\fboxsep}
}
\newenvironment{examplebox}{\par\begingroup
  \setlength{\fboxsep}{5pt}\findlength
  \setbox0=\vbox\bgroup\noindent
  \hsize=0.95\linewidth
  \begin{minipage}{0.95\linewidth}\normalsize}
    {\end{minipage}\egroup
    \textcolor{gray20}{\fboxsep1.5pt\fbox
     {\fboxsep5pt\colorbox{gray05}{\normalcolor\box0}}}
    \endgroup\par\noindent
    \normalcolor\ignorespacesafterend}

\newcounter{RQCounter}

\begin{document}

\title{Opportunities and Challenges in Code Search Tools}

\author{Chao Liu}
\email{liuchaoo@zju.edu.cn}
\affiliation{
  \institution{Zhejiang University, China}}

\author{Xin Xia}
\email{Xia@monash.edu}

\authornote{Corresponding Author: Xin Xia.}
\affiliation{
  \institution{Monash University, Australia}}
  
\author{David Lo}
\email{davidlo@smu.edu.sg}
\affiliation{
 \institution{Singapore Management University, Singapore}}
 
\author{Cuiyun Gao}
\email{gaocuiyun@hit.edu.cn}
\affiliation{
  \institution{Harbin Institute of Technology (Shenzhen), China}}
  
\author{Xiaohu Yang}
\email{yangxh@zju.edu.cn}
\affiliation{
  \institution{Zhejiang University, China}}
  
\author{John Grundy}
\email{John.Grundy@monash.edu}
\affiliation{
  \institution{Monash University, Australia}}

\renewcommand{\shortauthors}{Liu et al.}

\begin{abstract}

Code search is a core software engineering task. Effective code search tools can help developers substantially improve their software development efficiency and effectiveness. In recent years, many code search studies have leveraged different techniques, such as deep learning and information retrieval approaches, to retrieve expected code from a large-scale codebase. However, there is a lack of a comprehensive comparative summary of existing code search approaches. To understand the research trends in existing code search studies, we systematically reviewed 81 relevant studies. We investigated the publication trends of code search studies, analyzed key components, such as codebase, query, and modeling technique used to build code search tools, and classified existing tools into focusing on supporting seven different search tasks. Based on our findings, we identified a set of outstanding challenges in existing studies and a research roadmap for future code search research. 

\end{abstract}

\ccsdesc[500]{Software and its engineering~Search-based software engineering}

\keywords{code search, code retrieval, modeling}

\maketitle

\section{Introduction}\label{intro}

In modern software development, code search is one of the most frequent activities \cite{singer2010examination,sadowski2015developers,mcmillan2012recommending,xia2017developers}. Studies have shown that more than 90\% of developers' search efforts aim at finding code to reuse \cite{bajracharya2012analyzing}. This is because developers favor searching for existing or similar high-quality code to mitigate their learning burdens and enhance their software development productivity and quality \cite{brandt2009two,brandt2010example,gharehyazie2017some}. 

\textit{"Code search"} refers to retrieval of relevant code snippets from a code base, according to the intent of a developer that they have expressed as a search query \cite{li2013help,sim2011well,stolee2014solving,mcmillan2011portfolio,gharehyazie2017some,dit2013feature,rubin2013survey,kruger2019features}. With the advent of large code repositories and sophisticated search capabilities, code search is not only a key software development activity but also supports many other promising software engineering tasks. For example, code search tools are helpful for bug/defect localization \cite{nichols2010augmented,thummalapenta2009alattin,thummalapenta2011alattin,sisman2013assisting,lam2017bug,wang2020multi,akbar2019scor}, program repair \cite{afzal2019sosrepair,liu2019tbar}, and code synthesis \cite{poshyvanyk2009creating,raghothaman2016swim}, among others.


Despite the existence of numerous code search studies, to the best of our knowledge there has been no systematic study to summarize the key approaches and characteristics of existing code search tool research. Such a systematic review would help practitioners and researchers understand the current state-of-the-art code search tools and inspire their future studies. To perform a systematic review of this domain, we identified 81 relevant studies from three widely used electronic databases, including ACM Digital Library, IEEEXplore, and ISI Web of Science. We investigated the following research questions (RQs) in this study:


\begin{itemize}
    \item \textbf{RQ1.} \textit{What are the emerging publication trends for code search studies?} The reviewed 81 studies show that the popularity of the code search topic has increased substantially in recent years with a peak in 2019. 60\% of these studies were published in conference proceedings instead of journals. 83\% of studies contributed to this domain by proposing new tools rather than performing empirical/case studies of existing tools. \vspace{3pt}
    
    \item \textbf{RQ2.} \textit{What are the most important factors that contribute to existing code search tools?} From the 67 different code search tools reported, we found that deep learning (DL) based approaches are the most popular modeling techniques over the past two years. Code search tools can be classified into seven categories --  text-based code search, I/O example code search, API-based code search, code clone search, binary code search, UI search, and programming video search. Only 12 studies shared an accessible replication package link in their papers or provided their tool source code in GitHub.\vspace{3pt}
    
    \item \textbf{RQ3.} \textit{How do studies evaluate code search tools?} To evaluate a code search tool, most of the studies built codebases with method-level source code written in Java. they then performed code searches with free-form queries such as text and API names. Most studies manually checked the relevancy between a query and the returned code, and evaluated the tool performance in terms of popular ranking metrics, such as Precision and MRR (Mean Reciprocal Rank).
\end{itemize}

Based on these findings and the threats discussed in all reviewed code search studies, we observed a number of challenges in existing code search tools. Generally, the codebase scale used is limited with code written in only one programming language. The quantity of search queries is also limited and cannot cover developers' various search scenario in practical usage. The state-of-the-art tools based on learning models such as deep learning still cannot solve the code search problem very well. One major reason is that learning models are optimized with low quality and quantity of training data. The effectiveness of most code search tools is verified by use of manual evaluation that suffers from subjective bias. When measuring the tool performance, the search time and tool scalability are rarely assessed. Other important performance aspects, such as code diversity and conciseness, are not considered. These challenges provide some opportunities for further research studies:

\begin{figure}[t]
    \centering
    \includegraphics[width=0.7\linewidth]{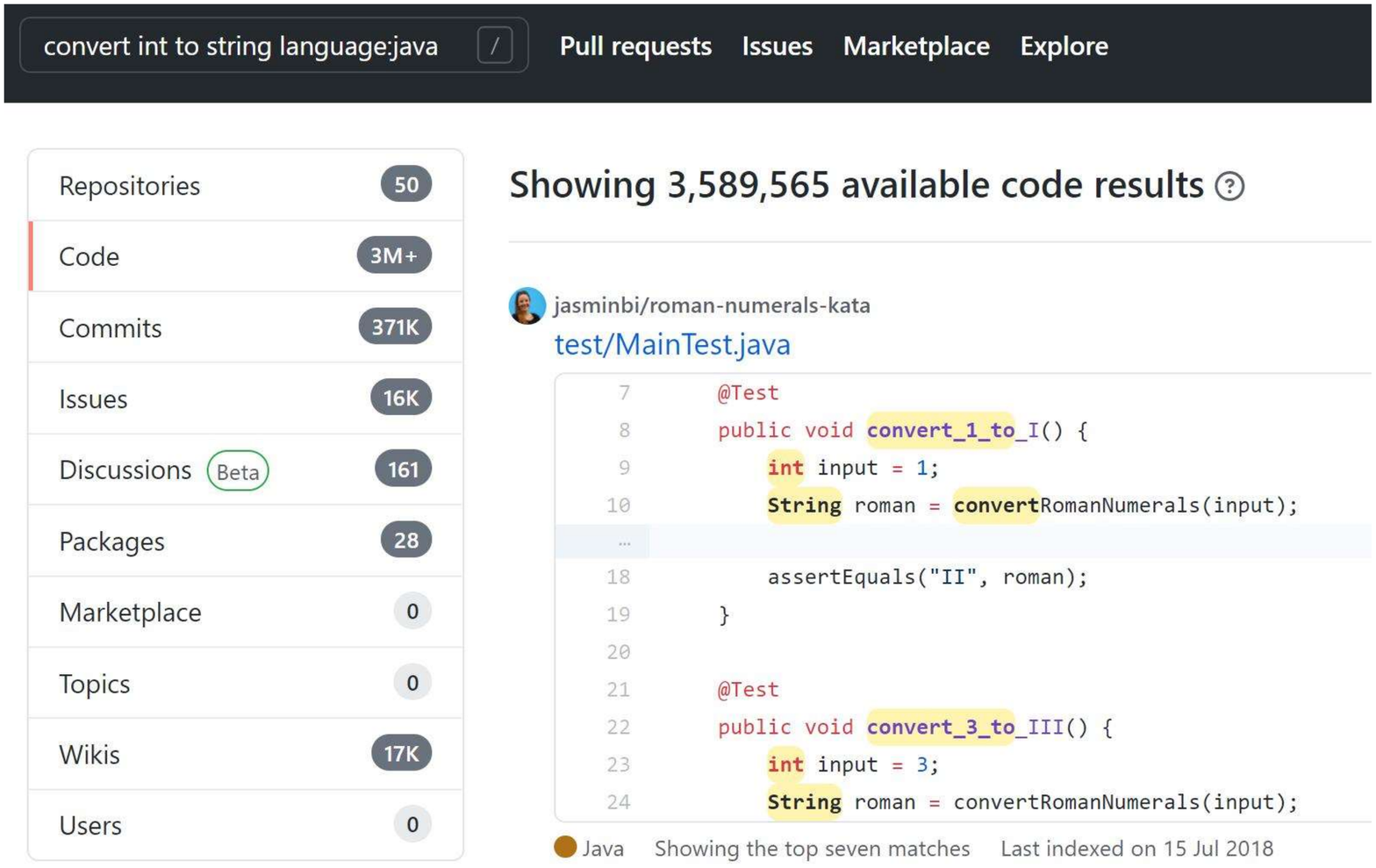}
    \caption{Searching Java code from GitHub with the text-based query "convert int to string".}
    \label{fig_github_example}
\end{figure}

\begin{itemize}
    \item \textbf{Benchmarks:} Developing a standard benchmark with large-scale code base written in multiple programming languages, various representative queries, and an automated evaluation method. \vspace{3pt}
    
    \item \textbf{Learning Models:} Improving the learning models with better quality of training data, code representation method, and loss functions for model optimization. \vspace{3pt}
    
    \item \textbf{Model Fusion:} Fusing different types of models -- such as deep learning based model, traditional IR-based model, and heuristic model -- to balance their advantages and disadvantages for further improvements. \vspace{3pt}
    
    \item \textbf{Cross-Language Searches:} Building a multi-language code search tool to mitigate the costly deployment of a tool for code in different programming languages. \vspace{3pt}
    
    \item \textbf{Search Tasks:} Supporting new kinds of code search tasks, such as searching UI (User Interface) code and the code used in programming videos. 
\end{itemize}

The main contributions of this study include:

\begin{itemize}
    \item A novel systematic review on 81 code search studies published until July 31, 2020 as a starting point for future research on code search.\vspace{3pt}
    
    \item Analysis of the fundamental components -- codebase, query, and model -- in 67 different code search tools to help researchers understand their characteristics.\vspace{3pt}
    
    \item Classification of code search tools into seven categories and analyzing the relationships between tools for each category as a basis for further comparisons and benchmarks.\vspace{3pt}
    
    \item Analysis of the outstanding opportunities and challenges in code search studies based on our findings to inspire further research in this area.\vspace{3pt}
    

\end{itemize}

The remainder of this paper is organized as follows. Section \ref{background} briefly introduces the usage of code search tools. Section \ref{method} presents our study methodology that we follow, and Sections \ref{rq1}-\ref{rq3} summarize the key research questions and their answers investigated in this study. Section \ref{challenge} discusses the challenges for the road ahead on code search studies and presents the potential research opportunities for future work. Section \ref{threat} shows the potential threats that may affect the validity of this review. Finally, Section \ref{conclude} provides a summary of this study.

\section{Background}\label{background}
The objective of code search tools is to retrieve relevant code from a large-scale codebase according to the intent of developers' search query. For example, GitHub search\footnote{https://github.com/search} is one type of tool widely used for searching for source code snippets from a large-scale codebase with millions of open source repositories. Fig. \ref{fig_github_example} shows an example of using the tool. After typing in the search query "convert int to string" with programming language choice "language:java", the GitHub search returns more than three million lines of potentially relevant code. However, the performance of this tool is not satisfactory, where the first returned code is not the expected code and it is time-consuming to check the relevancy of each code one by one. To improve such code search task performance, researchers have built many new tools. Some leverage DL techniques to improve the search accuracy, and some use clustering of returned code to reduce developers' efforts for code inspection.


The usual workflow when using such a code search tool involves the following six key components:

\point{Codebase.} In the code search task, the codebase defines the target search space, whose characteristics strongly affect the tool performance \cite{xia2017developers,sadowski2015developers,sim2011well}. Different code search studies and their tools build their codebases in different ways. For example, the codebase may be constructed by a set of source/compiled code written in different programming languages (e.g., Java, Python, and C/C++). The codebase scale also varies substantially and the code may be collected from various sources, such as GitHub, FDroid, and Stack Overflow. Section \ref{codebase} presents how developers build codebases from different perspectives.

\point{Query.} Code search tools take developers' queries as input. These queries reflect the developers' requirements during a specific software development task \cite{gu2018deep,cambronero2019deep,kim2018facoy}. Existing code search tools can support queries in different forms and this determines how developers use the code search tools. For example, free-form text written in natural language is the most common query, which is widely used for general search engines \cite{shuai2020improving,ye2020leveraging,linstead2009sourcerer}, such as GitHub search. Some code search tools support a more structured code-based query to find similar code in their codebase \cite{ragkhitwetsagul2019siamese,luan2019aroma,kamiya2002ccfinder}. A detailed analysis of query types is presented in Section \ref{query}.

\point{Model.} A code search tool should support the features of query and codebase. In general, researchers build code search tools by using three types of modeling techniques. The first is a traditional model that performs code search according to a relevancy algorithm (e.g., TF-IDF \cite{wu2008interpreting}) between query and candidate code \cite{lv2015codehow,mcmillan2013portfolio,nie2016query}. However, traditional models usually support only text-based queries. The second is a heuristic model that leverages the code analysis technique to capture the syntactic and semantic features in code and ranks code with customized matching approaches \cite{reiss2018seeking,zhong2009mapo,lee2010instant}. The final one is use of a learning model that learns the relationship between code and query by using a large-scale dataset. Most learning models embed query and code into a shared vector space so that their similarity can be measured by the cosine similarity. DL techniques have been widely used for building code search tools in recent years. This is because a DL model shows no limit on the types of query and codebase.  It learns features from large-scale data and this can substantially mitigate the difficulty in code understanding and representation as in the previous two model types \cite{ye2020leveraging,yao2019coacor,gu2018deep}. 

\point{Auxiliary Technique.} Although a learning model is promising compared with the traditional and heuristic models, we cannot say a learning model is superior to the other two. This is because code search models are usually associated with auxiliary techniques, such as query reformulation \cite{zou2020query,lv2015codehow}, code clustering \cite{liu2018supporting}, and learning from user feedback \cite{li2019reinforcement}. Using appropriate auxiliary techniques for a specific code search model can also improve their performance substantially \cite{liu2020simplifying}.

\point{Evaluation Method.} To evaluate the validity of a code search tool, the relevancy of the searched code list and a query should be assessed. Manual identification is the most prevalent method. This is because the ground-truth is difficult to measure \cite{gu2018deep,liu2020simplifying,shuai2020improving}. However, manual identification cannot scale to large numbers of queries. Therefore, researchers have also investigated other ways to mitigate manual efforts. Section \ref{evaluation} compares the tool evaluation methods that have been used.

\point{Performance.} Code search tool performance is based on its identified relevancy between query and the returned code. In most studies, code search tools care about the position of the correctly searched code in the result list. For example, MRR (mean reciprocal rank \cite{gu2018deep}) and NDCG (normalized discounted cumulative gain \cite{clarke2008novelty}) are two commonly used metrics. These metrics assume that developers prefer to find the "best" recommended code near the top of a result list \cite{shuai2020improving,yao2019coacor}. For example, supposing a tool returns three code with the expected one in the second and third places, the MRR metric measures the performance by the reciprocal rank of the first relevant code (i.e., 1/2). However, some developers want to search for more relevant code so that the ranking position can be ignored. Therefore, some code search studies evaluate the tool performance by using classification metrics, such as Precision, Recall, and F1-score (a harmonic average of Precision and Recall) \cite{chen2019capturing,lee2010instant}. For the above example with three returned code, the search precision equals to the number of relevant code divided by the total count (i.e., 2/3).

\section{Methodology}\label{method}
To perform a systematic review of code search tools, we followed the guidelines provided by Kitchenham and Charters \cite{keele2007guidelines} and Petersen et al. \cite{petersen2015guidelines}.


\subsection{Research Questions}\label{RQs}
We wanted to identify, summarize, classify, and analyze the empirical evidence concerning different code search studies published to date. To achieve this goal, we investigate three research questions (RQs): 

\begin{itemize}
    \item \textbf{RQ1.} \textit{What are the emerging publication trends for code search studies?} The goal of this RQ is to investigate the publication trends in terms of the publication year, publication venue, and contribution type (e.g., new tool and empirical study) of code search studies.\vspace{3pt}
    
    \item \textbf{RQ2.} \textit{What are the most important factors that contribute to existing code search tools?} This RQ investigates which modeling and auxiliary techniques have been used in different code search tools; how we can best classify these different tools; and how often do they provide accessible replication packages. \vspace{3pt}
    
    \item \textbf{RQ3.} \textit{How do studies evaluate code search tools?} This RQ aims to analyze four fundamental aspects for tool evaluation: codebase, query, evaluation method, and performance measures. \vspace{3pt}

\end{itemize}

Through analysis of these RQs we also identify limitations, gaps and future research recommendations from these studies. We use these to formulate our research roadmap for future code search studies.

\subsection{Search Strategy}

We identified a set of search terms in code search studies that were already known to us. We refined these search terms by checking the titles and abstracts of the relevant papers, combined them with logical "OR", and formed the search string: \emph{"code search" OR "code retrieval"}. We used the search string to perform an automated search on three widely used electronic databases including ACM Digital Library, IEEExplore, and ISI Web of Science. The search was performed on the title, abstract, and keywords of the papers. We conducted our search on July 31, 2020, and identified the studies published up until that date. As shown in Table \ref{tab_flow}, we retrieved 1,117 relevant studies with the automatic search from these three electronic databases. After discarding the duplicated studies, we obtained 692 code search studies.

\begin{table}[h]
    \centering
    \footnotesize
    \caption{Selection of code search studies.}
    \begin{tabular}{|lc|}
        \toprule
        \textbf{Process} & \textbf{\#Studies} \\
        \midrule
        ACM Digital Library & 165 \\
        IEEE Xplore & 322 \\
        Web of Science & 630 \\
        \midrule
        Automatic search from three electronic databases. & 1117\\
        Removing duplicated studies. & 692\\
        Excluding primary studies based on title and abstract. & 135\\
        Excluding primary studies based on full text - final left. & 81\\
        \bottomrule
    \end{tabular}
    \label{tab_flow}
\end{table}

\subsection{Study Selection}\label{criteria}

Once we retrieved the candidate studies relevant to the code search study, we performed a relevance assessment according to the following inclusion and exclusion criteria:\vspace{5pt}

\Checkmark \textit{The paper must be written in English.}\vspace{3pt}

\Checkmark \textit{The paper must involve at least one tool addressing the code search task.}\vspace{3pt}

\Checkmark \textit{The paper must be a peer-reviewed full research paper published in a conference proceeding or a journal.}\vspace{3pt}

\XSolidBrush \textit{Keynote records and grey literature are excluded.}\vspace{3pt}

\XSolidBrush \textit{Conference studies with extended journal versions are discarded.}\vspace{3pt}

\XSolidBrush \textit{The studies that propose new code search tools but did not evaluate their performance are excluded.}\vspace{3pt}

\XSolidBrush \textit{The studies that apply existing code search tools for other software engineering tasks (e.g., bug localization and program repair) are ruled out.}\vspace{5pt}

The inclusion and exclusion criteria were piloted by the first and forth authors starting with the assessment of 30 randomly selected primary studies. The reliability of the inclusion/exclusion decisions was measured using pairwise inter-rater reliability with Cohen's Kappa statistic \cite{cohen1960coefficient}. The agreement rate in the pilot study was "moderate" (0.59). The pilot study helped us to develop a collective understanding of the inclusion/exclusion criteria. Then, an assessment was performed for the full list of the identified studies. The agreement rate in the full assessment was "substantial" (0.73). Disagreements were resolved after open discussions between first and fourth authors. For any case that they did not reach a consensus, the third author was consulted as a tie-breaker. Specifically, we took two weeks to finish the study selection process. As shown in Table \ref{tab_flow}, we identified 135 code search studies by inspecting the title and abstract of the retrieved studies. After checking the full text of the remaining studies, we finally obtained 81 relevant code search studies.

\subsection{Data Extraction}

To answer the three research questions above, we read the 81 papers carefully and extracted the required data as summarized in Table \ref{tab_data}. Our data collection mainly focused on four kinds of information: publication information, study background, tool details, and experimental setup. To suppress the effect of subjective bias, the data collection was performed by the first and fourth authors, and verified by two senior PhD students who are not co-authors of this study and majored in computer science. 

\begin{table}[h]
    \centering
    \footnotesize
    \caption{Extracted data for research questions.}
    \begin{tabular}{|llL{0.7\linewidth}|}
        \toprule
        \textbf{RQ} & \textbf{Description} & \textbf{Extracted Study Data}\\
        \midrule
        RQ1 & Publication Trend & Publication year, publication venue, publication type (i.e., new tool, empirical study, and case study). \\
        RQ2 & Modeling Techniques & Model (type, major technique), auxiliary technique, tool descriptions (background, motivation, application scenario, baseline tools), Replication package link.\\
        RQ3 & Evaluation Components & Codebase (type, granularity, language, scale, source), query (type, scale, source), evaluation method, performance measure.\\
        \bottomrule
    \end{tabular}
    \label{tab_data}
\end{table}

\section{RQ1: What are the emerging publication trends for code search studies?}\label{rq1}

We analyze the publication information from the 81 code search studies retrieved from Section \ref{method}, and discuss the key emerging publication trends.

\subsection{Publication Trend}

\begin{figure}[h]
    \centering
    \small
    \subfigure[Number of publications per year.]{
    \begin{tikzpicture}
    \begin{axis}[
        ybar, 
        label style={font=\small}, 
        tick label style={font=\small},
        width = \linewidth,
        height=4cm,
        ymax=23,
        xmin=2001,
        xmax=2021,
        enlargelimits=0,
        ylabel={\#Publications},
        y label style={at={(0.05,0.5)}},
        symbolic x coords={2001,2002,2003,2004,2005,2006,2007,2008,2009,2010,2011,2012,2013,2014,2015,2016,2017,2018,2019,2020,2021}, 
        xtick=data, 
        nodes near coords, 
        nodes near coords align={vertical},]
        \addplot[black, fill=gray] coordinates {(2002,1)(2003,0)(2004,0)(2005,0) (2006,0)(2007,1)(2008,0)(2009,5)(2010,4)(2011,3)(2012,1)(2013,3)(2014,6)(2015,7)(2016,9)(2017,4)(2018,11)(2019,18)(2020,8)};
    \end{axis}
    \end{tikzpicture}}
    \subfigure[Cumulative number of publications per year, which shows an increasing trend ($p$-value<0.01) tested by the Cox Stuart trend test \cite{cox1955some} at a 5\% significance level.]{
    \begin{tikzpicture}
    \begin{axis}[
        ybar, 
        label style={font=\small}, 
        tick label style={font=\small},
        width = \linewidth,
        height=4cm,
        ymax=110,
        xmin=2001,
        xmax=2021,
        enlargelimits=0,
        ylabel={\#Publications},
        y label style={at={(0.05,0.5)}},
        symbolic x coords={2001,2002,2003,2004,2005,2006,2007,2008,2009,2010,2011,2012,2013,2014,2015,2016,2017,2018,2019,2020,2021}, 
        xtick=data, 
        nodes near coords, 
        nodes near coords align={vertical},]
        \addplot[black, fill=gray] coordinates {(2002,1)(2003,1)(2004,1)(2005,1) (2006,1)(2007,2)(2008,2)(2009,7)(2010,11)(2011,14)(2012,15)(2013,18)(2014,24)(2015,31)(2016,40)(2017,44)(2018,55)(2019,73)(2020,81)};
    \end{axis}
    \end{tikzpicture}}
    \caption{Publication trend in years.}
    \label{fig_publication_trend}
\end{figure}
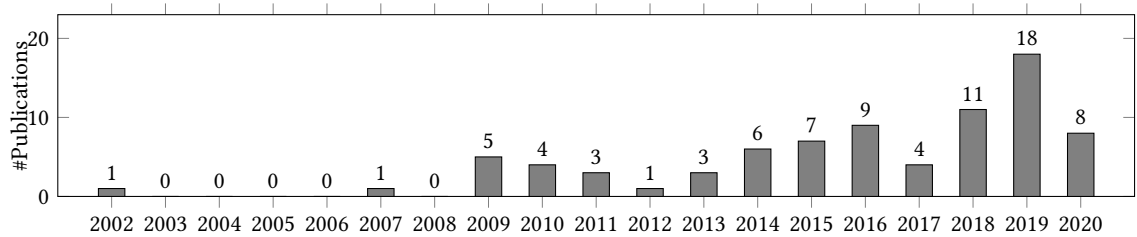
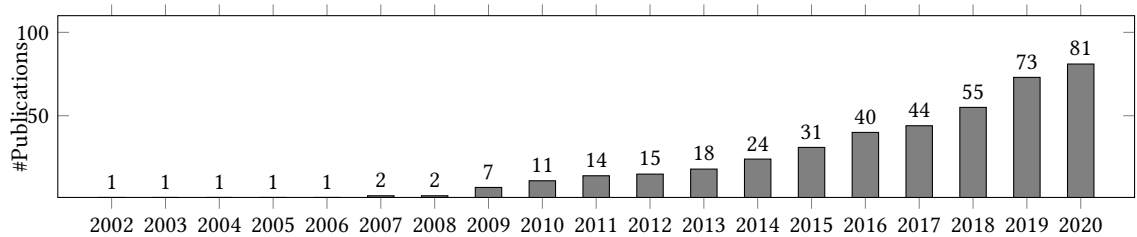

Fig. \ref{fig_publication_trend}(a) shows the number of studies that were published each year. We can see that the first code search study we found was published in 2002. The popularity of code search studies has gradually been increasing from 2009, and the publication peak occurs in 2019 with 22.2\% of the total numbers (and not all 2020 papers have yet appeared of course). Fig. \ref{fig_publication_trend}(b) illustrates the cumulative counts of numbers shown in Fig. \ref{fig_publication_trend}(a). To test the trend (i.e., increasing, decreasing, or neither) of the cumulative publication number, we performed a Cox Stuart trend test \cite{cox1955some} at a 5\% significance level. The statistical result shows a substantially increasing trend with $p$-value<0.01, which implies the growing popularity of the code search study in the last 18 years.

\subsection{Publication Venues and Contribution Types}

\begin{figure}[h]
    \centering
    \small
    \subfigure[Publication venue types.]{
    \begin{tikzpicture}[scale=0.5]
    \pie[rotate=330, text=pin, color={gray!80, gray!40, gray!10}]{60/Conference (49), 40/Journal (32)}
    \end{tikzpicture}}
    \subfigure[Contribution types.]{
    \begin{tikzpicture}[scale=0.5]
    \pie[rotate=290, text=pin, color={gray!80, gray!40, gray!10}]{83/New Tool (67), 12/Empirical Study (10), 5/Case Study (4)}
    \end{tikzpicture}}
    \caption{Publication Venues and Contribution Types}
    \label{fig_veune_type}
\end{figure}
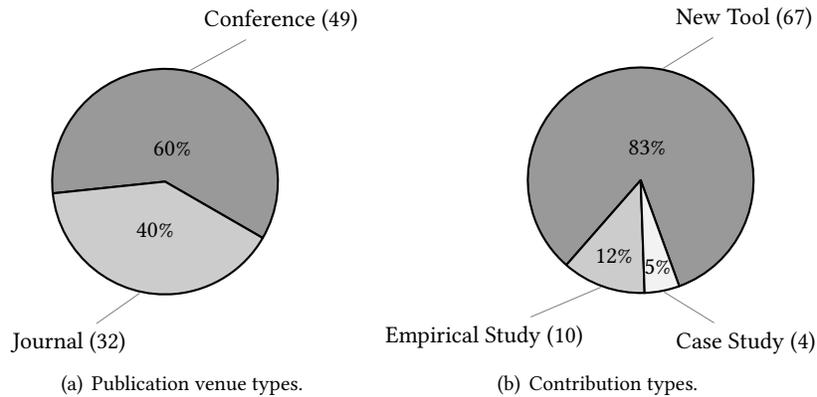

The 81 reviewed studies were published in various conference proceedings and journals. Fig. \ref{fig_veune_type}(a) shows that 60\% of them were published in conference proceedings. Fig. \ref{fig_veune_type}(b) illustrates how studies contributed to the code search task. We can notice that 83\% of the studies proposed new tools, while 12\% of the studies performed empirical studies to analyze the historic data of existing code search tools \cite{bajracharya2009mining,ratanotayanon2010my,bajracharya2012analyzing,ge2014developers,martie2015sameness,gharehyazie2017some,xia2017developers,rahman2018evaluating,yan2020code,damevski2016field}. The remaining 5\% studies performed case studies in real world, especially for enterprise usage, to investigate developers' experience and expectations of code search tools \cite{sadowski2015developers,damevski2014case,sim2011well,starke2009searching}. 

Table \ref{tab_venue_type} lists the top publication venues with at least two code search studies. These venues include a total of 56 studies, 69.1\% of the total reviewed studies. These publication venues publish various kinds of code search studies: studies that propose new tools (46), empirical studies (9), case study (1). We can also observe that among these 17 venues, the top-5 popular conferences these works were published are MSR, ICSE, ASE, FSE, and EMSE; meanwhile, the top-5 journals are TSE, TOSEM, SPE, ASEJ, and TSC.


\begin{table}[h]
    \centering
    \footnotesize
    \caption{Top publication venues with at least two code search studies (NT: new tool; ES: empirical study; CS: case study; TS: Total Studies).}
    \setlength{\tabcolsep}{3.4pt}{
    \begin{tabular}{|cL{0.69\linewidth}cccc|}
        \toprule
        \textbf{Short Name} & \textbf{Full Name} & \textbf{NT} & \textbf{ES} & \textbf{CS} & \textbf{TS}  \\
        \midrule
        MSR & International Conference on Mining Software Repositories & 3 & 4 & 0 & 7 \\
        ICSE & International Conference on Software Engineering &6&0&0& 6\\
        ASE & International Conference Automated Software Engineering &6&0&0& 6\\
        FSE & Symposium on the Foundation of Software Engineering & 4 & 1 & 0 & 5\\
        TSE & Transactions on Software Engineering &4&0&0& 4\\ EMSE & International Symposium on Empirical Software Engineering and Measurement &1&2&0& 3\\   
        TOSEM & Transactions on Software Engineering and Methodology &2&0&1& 3\\
        SANER & International Conference on Software Analysis, Evolution, and Reengineering &2&1&0& 3\\
        SPE & Software-Practice \& Experience &3&0&0& 3\\
        ICPC & International Conference on Program Comprehension &1&1&0& 2\\
        ASEJ & Automated Software Engineering &2&0&0& 2\\
        TSC & Transactions on Service Computing &2&0&0& 2\\
        JSS & Journal of Systems and Software &2&0&0& 2\\
        APSEC & Asia-Pacific Software Engineering Conference &2&0&0& 2\\
        WWW & The World Wide Web Conference &2&0&0& 2\\
        MLPL & International Workshop on Machine Learning and Programming Languages &2&0&0& 2\\
        Access & IEEE Access &2&0&0& 2\\
         - & \textbf{Total} & \textbf{46} & \textbf{9} & \textbf{1} & \textbf{56} \\
        \bottomrule
    \end{tabular}}
    \label{tab_venue_type}
\end{table}

\begin{framed}
\filbreak
\noindent\textit{\textbf{Summary of answers to RQ1:}}
\begin{itemize}
\item \textit{Code search started to be considered in the software engineering research literature in 2002, and its popularity continues to increase with a current 
peak so far in 2019.}\vspace{3pt}
\item \textit{60\% of the studies are published in conferences (rather than journals).}\vspace{3pt}
\item \textit{67 of the studies propose new code search tools.}
\end{itemize}
\end{framed}

\section{RQ2: What are the most important factors that contribute to existing code search tools?}\label{rq2}
We first present an analysis of the modeling and auxiliary techniques used by the reviewed code search tools in Section \ref{model} and \ref{technique} respectively. Section \ref{classify} then presents a classification of the tools into seven categories and describes how these tools work in general. Section \ref{replication} investigates how often code search studies provide accessible replication packages.

\subsection{Models}\label{model}

For a given query and a codebase, the objective of a code search model is to correctly measure the semantic relevancy between the query and candidate code snippets in the codebase, and retrieve the top-k code according to their relevancy scores. Table \ref{tab_model} shows the main models used in the reviewed code search studies. TF-IDF (Term Frequency-Inverse Document Frequency), BM25 (Best Match 25), and deep learning are the most frequently used modeling techniques in the last three years. To analyze their general features, we classified them into four categories. This includes traditional models (regarding code as text and searching code with IR-based techniques), learning models (leveraging probabilistic model or neural network to learn the relationship between query and code), heuristic models measuring the semantic similarity between query and code by using designed features, and online models (existing online code search tools, e.g., GitHub search).  

\begin{table}[h]
    \centering
    \footnotesize
    \caption{Number of code search models studied in each year from 2002 to 2020 .}
    \setlength{\tabcolsep}{5.3pt}{
    \begin{tabular}{|llccccccccccccccc|}
        \toprule
        \textbf{Type} & \textbf{Model} & \textbf{02}  &\textbf{07} & \textbf{09} & \textbf{10} & \textbf{11} & \textbf{12} & \textbf{13} & \textbf{14}& \textbf{15}& \textbf{16}& \textbf{17}& \textbf{18}& \textbf{19}& \textbf{20} & \textbf{Total} \\
        \midrule
        \multirow{4}{*}{Traditional}
        & TF-IDF &-&-&2&-&-&-&2&-&1&3&1&3&3&2&17\\
        & BM25 &-&-&-&-&-&-&-&-&-&-&-&2&5&1&8\\
        & EBM &-&-&-&-&-&-&-&-&1&-&1&-&-&-&2\\ 
        & SSI &-&-&-&1&-&-&-&-&-&-&-&-&-&-&1\\
        \midrule
        \multirow{2}{*}{Heuristic}
        & Customized Matching &1&-&1&1&1&-&1&1&2&-&-&3&2&-&13\\
        & Graph Search &-&-&-&-&-&-&-&-&-&2&-&-&-&-&2\\
        \midrule
        \multirow{3}{*}{Learning}
        & Deep Learning &-&-&-&-&-&-&-&-&-&2&-&1&4&3&10\\
        & fastText &-&-&-&-&-&-&-&-&-&-&-&1&2&-& 3\\
        & Probabilistic Model &-&-&-&-&-&-&-&-&1&-&-&-&-&-&1\\
        \midrule
        \multirow{3}{*}{Online}
        & GitHub Search &-&-&-&-&-&-&-&-&-&1&-&-&2&1&4\\
        & Google Code &-&1&-&1&-&-&-&1&-&-&-&-&-&-&3\\
        & Sourcerer &-&-&-&-&1&-&-&2&-&-&-&-&-&-&3\\
        \midrule
        -& \textbf{Total} & 1 & 1 &3&3&2&0&3&4&5&8&2&10&18&7&67\\
        \bottomrule
    \end{tabular}}
    \label{tab_model}
\end{table}

\vspace{5pt}\noindent\textbf{Traditional Models.}
Table \ref{tab_model} shows that 24 code search models leveraged the traditional ranking algorithms TF-IDF \cite{wu2008interpreting} and BM25 \cite{robertson2009probabilistic} to measure the relevancy between a query and a candidate code based on the frequency of their shared words \cite{nie2016query,lv2015codehow}. TF-IDF is a simple and effective ranking method. A high TF-IDF score indicates that the query words frequently appeared in the relevant code but rarely occurred in other irrelevant code \cite{bao2020psc2code,martie2015codeexchange}. BM25 is an improvement of TF-IDF, which restricts the effect of the terms with unexpectedly high frequency under a limited upper bound and balances the term importance for code with different sizes \cite{robertson2009probabilistic}. 

However, the above models only connect query words with the Boolean operator "OR" implicitly and cannot address the "AND" operation. To support code search with complete Boolean queries, two studies \cite{yang2017iecs,lv2015codehow} in Table \ref{tab_model} extended query with related words and the operator "AND", and leveraged the Extended Boolean Model (EBM) \cite{salton1983extended} to calculate the relevancy between query and code. We can notice that all the above models match the words in query and code directly. However, a search query can be described using many different words, and the code may also express a requirement completely different from the search intent. In this case, the direct word matching is prone to failure. To overcome this issue, the Structural Semantic Indexing (SSI) is a traditional choice \cite{rosario2000latent}. SSI represents the codebase as a word-code matrix that records the frequency of a term that occurred at each code. The matrix is reduced via Singular Value Decomposition (SVD) to filter out the noise found in a code so that two code which have the same semantics are located close to one another in a multi-dimensional space \cite{bajracharya2010leveraging}. However, a major limit is that the query should be one of the codes in the codebase.

\vspace{5pt}\noindent\textbf{Heuristic Models.} Traditional models mainly regard code as text. But code is not just text -- it is written in highly structured programming languages with specific keywords, syntactic rules, and semantic representations and meanings \cite{gu2018deep,zhang2019novel}. To better express code semantics, Table \ref{tab_model} shows that 15 code search studies developed heuristic models based on researchers' domain knowledge to search code in a more intuitive way. Graph search is one representative method, which represents code as a control flow graph \cite{ding2016kam1n0} or a call graph \cite{li2016relationship}. The code search is then transformed into a sub-graph matching issue between query and code. However, most heuristic models search code by designing a customized Matching between query and code \cite{xie2019user,xue2018accurate,balachandran2015query}, such as designed similarity score \cite{xie2019user,behrang2018guifetch}, static analysis \cite{thummalapenta2007parseweb,reiss2009semantics}, and dynamic analysis \cite{stolee2016code,chen2020enhancing}.

\vspace{5pt}\noindent\textbf{Learning Models.} The semantic gap between query and code is the major challenging issue for traditional and heuristic code search models. To address this challenge, researchers have built learning models that capture the correlation between query and code from large-scale training data. To reduce the relevancy ranking problem to an application of the probability theory, a probabilistic model \cite{allamanis2015bimodal} computes the relevancy score between a query and a code as the probability that the code will be relevant to the query. This reduces the relevancy ranking problem to an application of the probability theory. 

Many other learning models leverage techniques to embed query and code into a shared vector space. The code search problem can then be performed by measuring the cosine similarity between vectors. Table \ref{tab_model} shows that three code search studies adopted the fastText \cite{mikolov2013efficient}, a well-known embedding library based on a shallow neural network succeeded in text classification, for query/code embedding. However, we can see that more code search studies preferred to use deep learning techniques for embedding, including CNN (convolutional neural network) \cite{shuai2020improving,huang2020code}, LSTM (long short-term memory) \cite{ye2020leveraging,yao2019coacor,wan2019multi,gu2018deep,iyer2016summarizing,chen2019capturing,white2016deep}, GGNN (gated graph neural network) \cite{wan2019multi}, and FFNN (feed-forward neural network) \cite{ding2019asm2vec}.

\vspace{5pt}\noindent\textbf{Online Search Models.} Table \ref{tab_model} shows that ten code search studies built models based on online search engines, including GitHub Search\footnote{https://github.com/search}, Google Code search, and Sourcerer. These studies focus on how to refine the search results returned by these online search engines. 

\subsection{Auxiliary Technique}\label{technique}

Simply building code search tools with techniques listed in Table \ref{tab_model} may not be enough for practical usage scenarios. Therefore, researchers have also utilized auxiliary techniques to improve the search effectiveness and efficiency. Table \ref{tab_auxiliary} presents the auxiliary techniques used in different years. 

\begin{table}[h]
    \centering
    \footnotesize
    \caption{Number of studies used auxiliary techniques in each year from 2002 to 2020.}
    \setlength{\tabcolsep}{6pt}{
    \begin{tabular}{|lccccccccccccccc|}
        \toprule
        \textbf{Technique} & \textbf{02}  &\textbf{07} & \textbf{09} & \textbf{10} & \textbf{11} & \textbf{12} & \textbf{13} & \textbf{14}& \textbf{15}& \textbf{16}& \textbf{17}& \textbf{18}& \textbf{19}& \textbf{20} & \textbf{Total} \\
        \midrule
        Inverted Index &-&-&2&1&-&-&3&-&1&3&1&3&6&2& 22\\
        Query Reformulation &-&-&-&-&-&-&-&2&2&1&2&3&6&2& 18\\
        Clustering &-&-&1&1&1&-&1&1&1&-&-&2&3&-&11\\        Learning from User Feedback &-&-&-&-&-&-&-&-&-&-&-&-&1&-&1\\
        \textbf{Total} &0&0&2&2&1&0&4&4&4&4&3&8&16&4&52\\
        \bottomrule
    \end{tabular}}
    \label{tab_auxiliary}
\end{table}

\vspace{5pt}\noindent\textbf{Inverted Index.} 
Time efficiency is of high importance in code search due to the need to search a large-scale codebase. To accelerate the search response time, 22 code search studies indexed code by leveraging the Lucene tool \cite{mccandless2010lucene}. Lucene is an efficient text-based search engine, which divides indexing information for any given term into blocks, and builds a parallel structure called a skip list \cite{zhang2019s3} to allow queries to efficiently jump over a set of code that does not match a query. One study also applied R*tree instead of Lucene \cite{lee2010instant}, a multi-dimensional structure for vector indexing \cite{beckmann1990r}.

\vspace{5pt}\noindent\textbf{Query Reformulation.} 
A free-form text query written in natural language is usually short and misses related contexts. Thus, a code search tool likely returns many irrelevant code snippets for its queries without complete and precise semantics. Therefore, many tools have reformulated developers' queries before performing the code search. This includes techniques such as expanding queries with related words from Stack Overflow \cite{zou2020query,sirres2018augmenting}, extending query words with relevant APIs or class names \cite{lv2015codehow,zhang2017expanding}, replacing query words with better synonyms in codebase \cite{lemos2014thesaurusExpand,lemos2014thesaurus}.

\vspace{5pt}\noindent\textbf{Clustering.} 
It is time-consuming for developers to inspect each code snippet returned by a tool one by one. Therefore, researchers have tried to reorganize the results list by clustering similar code snippets \cite{liu2018supporting,keivanloo2014spotting,gu2019codekernel,luan2019aroma,kim2018facoy}. In this case, much developer effort can be saved by only checking the representative code examples. If one developer is interested in one representative, they can check the corresponding cluster later. However, Table \ref{tab_auxiliary} shows that only 11 of the reviewed code search tools actually improved the code search results by using such a clustering technique. Thus, it is suggested for further studies to consider results clustering improvement as an important tool component. 

\vspace{5pt}\noindent\textbf{Learning from User Feedback.} 
After a search tool returns a list of relevant codes, developers usually check each code one by one and inspect the relevant ones. Developers' feedback on search relevancy can help a tool to identify users' real interests and continuously optimize the tool performance. To reach this goal, researchers have leveraged reinforcement learning to capture developers' preferences \cite{li2019reinforcement}. However, it is not easy to obtain such user feedback. This may be the reason why researchers have rarely investigated incorporation of feedback learning into code search tools. This is another promising area for further research.

\subsection{Classification of Code Search Tasks}\label{classify}
To understand how different code search tools work, the reviewed 67 code search tools and classified them into seven categories, as shown in Table \ref{tab_tasks}: \textit{1) Text-based code search} -- searches source code shared with the same semantics as developers' text-based search queries; \textit{2) Code clone search} -- uses source code as input and finds similar code from a codebase; \textit{3) I/O example code search} -- aims to find code that matches a given input/output example; \textit{4) API-based code search} -- finds representative API examples from a codebase according to a given API name; \textit{5) Binary code search} -- is similar to the code clone search task but focuses on the binary code (i.e., the compiled source code); \textit{6) UI code search} -- retrieves UI implementation code that matches developers' manually sketched UI images; \textit{7) Programming video search} -- is a special variant of the text-based code search task but searches code in programming videos. 

Table \ref{tab_task_distribution} shows the number of studies published from 2002 to 2020 for each of these code search task classifications. We can see that text-based code search is the most popular task with a total of 34 proposed tools and it is also the most frequently investigated task in the recent three years. Moreover, UI code search and the programming video search are two emerging tasks, which require more attention from further studies. 

\begin{table}[h]
    \centering
    \footnotesize
    \caption{Classification of code search tasks.}
    \setlength{\tabcolsep}{11pt}{
    \begin{tabular}{|cccccc|}
        \toprule
        \textbf{No.} & \textbf{Code Search Task} & \textbf{Query} & \textbf{Codebases} & \textbf{\#Studies} & \textbf{Percent}\\
        \midrule
        1 & Text-Based Code Search & Text & Source Code & 34 & 51\%\\
        2 & Code Clone Search & Source Code & Source Code & 9 & 14\% \\
        3 & I/O Example Code Search & Input/Output Example & Source Code & 8& 12\%\\
        4 & API-Based Code Search & API & Source Code & 7&11\%\\
        5 & Binary Clone Search & Binary Code & Binary Code & 5&7\%\\
        6 & UI Code Search & UI Sketch & UI Code & 3&4\%\\
        7 & Programming Video Search & Text & Code in Video & 1&1\%\\
        -& \textbf{Total} &-&-&67&100\%\\
        \bottomrule
    \end{tabular}}
    \label{tab_tasks}
\end{table}

\begin{table}[h]
    \centering
    \footnotesize
    \caption{Number of code search tasks studied in each year ranging from 2002 to 2020.}
    \setlength{\tabcolsep}{6.1pt}{
    \begin{tabular}{|lccccccccccccccc|}
        \toprule
        \textbf{Task} & \textbf{02} &\textbf{07} & \textbf{09} & \textbf{10} & \textbf{11} & \textbf{12} & \textbf{13} & \textbf{14}& \textbf{15}& \textbf{16}& \textbf{17}& \textbf{18}& \textbf{19}& \textbf{20} & \textbf{Total} \\
        \midrule
        Test-Based Code Search &-&-&1&1&-&-&1&1&3&4&2&6&10&6&34\\
        Code Clone Search &1&-&-&1&-&-&-&-&1&2&-&1&3&-&9\\ 
        I/O example Code Search &-&1&1&-&1&-&-&2&-&1&-&-&1&1&8\\
        API-Based Code Search &-&-&1&1&1&-&1&-&1&-&-&-&2&-&7\\
        Binary Code Search &-&-&-&-&-&-&1&1&-&1&-&1&1&-&5\\
        UI Code Search &-&-&-&-&-&-&-&-&-&-&-&2&1&-&3\\
        Programming Video Search &-&-&-&-&-&-&-&-&-&-&-&-&-&1&1\\
        \textbf{Total} & 1& 1&3&3&2&0&3&4&5&8&2&10&18&7&67\\
        \bottomrule
    \end{tabular}}
    \label{tab_task_distribution}
\end{table}

\subsubsection{Text-Based Code Search}\label{text_based_code_search}

The goal of text-based code search is to retrieve source code from a large-scale code corpus that most closely matches the free-form input by developers \cite{cambronero2019deep,bao2020psc2code}. Reusing existing code can largely boost developers' coding efficiency and potentially also quality, by reusing high quality code examples. Text-based code search is frequently studied because this task aims to improve the performance of frequently used code search engines in practice, such as GitHub search. Table \ref{tab_text_based_code_search} presents text-based code search tools with their major and auxiliary modeling techniques. This table also shows how different tools are related to each other in terms of their compared baseline tools.

\begin{table}
    \centering
    \footnotesize
    \caption{Text-based code search tools.}
    \setlength{\tabcolsep}{15pt}{
    \begin{tabular}{|cccc|}
        \toprule
        \textbf{No.} & \textbf{Year} & \textbf{Tool Name} & \textbf{Baselines} \\
        \midrule
        1&2020 & CARLCS-CNN \cite{shuai2020improving} & DeepCS \\
        2&2020 & QESC2 \cite{zou2020query} & QECK, CodeHow\\
        3&2020 & Ye20 \cite{ye2020leveraging} & DeepCS, CoaCor\\
        4&2020 & CDRL \cite{huang2020code} & DeepCS, QECK, CodeHow\\
        5&2020 & CodeMF \cite{hu2020unsupervised} & QECK\\
        6&2019 & CoaCor \cite{yao2019coacor} & DeepCS, CODE-NN\\
        7&2019 & Wu19 \cite{wu2019code} & CodeHow\\
        8&2019 & MMAN \cite{wan2019multi} & CodeHow, DeepCS\\
        9&2019 & NQE \cite{liu2019neural} & NCS\\
        10&2019 & Cosoch \cite{li2019reinforcement} & -\\
        11&2019 & QESC1 \cite{huang2019deep} & QECK, CodeHow\\
        12&2019 & QREC \cite{huang2019enhance} & QECK, CodeHow\\
        13&2019 & UNIF \cite{cambronero2019deep} & NCS, DeepCS\\
        14&2019 & GKSR \cite{huang2019qe} & QECK, CodeHow\\
        15&2019 & QESR \cite{jin2019query} & QECK, CodeHow\\
        16&2018 & GitSearch \cite{sirres2018augmenting} & -\\
        17&2018 & NCS \cite{sachdev2018retrieval} & -\\
        18&2018 & CodeNuance \cite{liu2018supporting} & CodeExchange\\
        19&2018 & QECC \cite{huang2018query} & QECK, CodeHow\\
        20&2018 & DeepCS \cite{gu2018deep} & CodeHow, Sourcerer\\
        21&2018 & Codepus \cite{lee2018comment} & -\\
        22&2017 & Zhang17 \cite{zhang2017expanding} & -\\
        23&2017 & SnippetGen \cite{yang2017iecs} & CodeHow\\
        24&2016 & QECK \cite{nie2016query} & Portfolio, Keivanloo14\\
        25&2016 & RACKS \cite{li2016relationship} & -\\
        26&2016 & ROSF \cite{jiang2016rosf} & Portfolio, Keivanloo14\\
        27&2016 & CODE-NN \cite{iyer2016summarizing}& -\\
        28&2015 & CodeExchange \cite{martie2015codeexchange}& -\\
        29&2015 & CodeHow \cite{lv2015codehow}& Sourcerer\\
        30&2015 & Allamanis15 \cite{allamanis2015bimodal}& -\\
        31&2014 & Keivanloo14 \cite{keivanloo2014spotting}& -\\
        32&2013 & Portfolio \cite{mcmillan2013portfolio}& -\\
        33&2010 & Bajracharya10 \cite{bajracharya2010leveraging}& Sourcerer\\
        34&2009 & Sourcerer \cite{linstead2009sourcerer}& -\\
        \bottomrule
    \end{tabular}}
    \label{tab_text_based_code_search}
\end{table}

\vspace{5pt}\noindent\textbf{Traditional Models.} 
Sourcerer \cite{linstead2009sourcerer} is one of the simplest tools. It simply regards code as plain text and ranks candidate found code in the codebase by the classic information retrieval approach TF-IDF (term frequency-inverse document frequency) \cite{haiduc2013automatic}. A higher TF-IDF score means that the query words frequently appeared in the relevant code but rarely occurred in other irrelevant code. To accelerate the search response time, Sourcerer leveraged the Lucene library \cite{mccandless2010lucene} to build an inverted index for the large-scale codebase. Finally, for a given query, Sourcerer re-ranks the candidate code according to their popularity within the code dependency graph \cite{wilde1989dependency}.

To improve code search quality, researchers have tried various approaches. Martie et al. \cite{martie2015codeexchange} provided code searchers with more advanced choices (e.g., package, class, method, parameters, etc.). Jiang et al. \cite{jiang2016rosf} refined the result list by building a supervised ranker to predict the relevancy of a candidate code to a query. Li et al. \cite{li2019reinforcement}, and leveraged reinforcement learning techniques to learn fromn user feedback -- whether a code snippet found is relevant to a search query or not -- from developers' search sessions. To mitigate developers' inspection efforts, researchers used code clone detection methods \cite{roy2007survey,sheneamer2016survey} to cluster candidate code and present the representative ones for developers \cite{keivanloo2014spotting,liu2018supporting}. They have also augmented searched code with contextual information in terms of a chain of methods in the code dependency graph \cite{mcmillan2013portfolio,li2016relationship}.

However, the major challenge is the semantic gap between query and code. Code is not like a search query and code is written in a highly structured programming language with different syntactic rules and semantic representation \cite{lv2015codehow,zou2020query,hu2020unsupervised}. To address this issue, researchers have proposed many query processing techniques to align this semantic gap, such as replacing query words with appropriate synonyms that occur in the codebase \cite{wu2019code}; and expanding query words with code changes (e.g., pull requests and commits) in the development history \cite{huang2018query,jin2019query,huang2019qe,jin2019query,huang2019enhance,huang2019deep,zou2020query}. It has been observed that APIs are an important factor to complement the missing semantics in queries \cite{bajracharya2010leveraging}. Researchers have thus expanded query words with relevant APIs or class names from official API documents \cite{lv2015codehow}, codebases \cite{yang2017iecs}, or Stack Overflow posts \cite{barua2014developers,nie2016query,hu2020unsupervised,zhang2017expanding,sirres2018augmenting}.

\vspace{5pt}\noindent\textbf{Learning Models.}
To incorporate more domain knowledge into code search tools, researchers have built various learning models. Allamanis et al. \cite{allamanis2015bimodal} explained the code search task as a probabilistic model, namely the probability that a code would be retrieved to match the input query. Other proposed tools score query-code pairs by a trained multiplicative model \cite{mitchell2010composition}. This shows that the query and code can be jointly modelled, inspiring other new learning models. Iyer et al. \cite{iyer2016summarizing} proposed CODE-NN that leverages LSTM to build a translation model from query to code. To train the model, CODE-NN collected posts from Stack Overflow, where the question and corresponding code in the post are used as the training data. 

To learn a better representation of code and query, Gu et al. \cite{gu2018deep} proposed the tool DeepCS that represents code by separate components, including method name, API sequence, and word set in the method body. Their tool embeds query and code into vectors so that code search can be performed by measuring the cosine similarity between vectors. DeepCS is trained by the code and corresponding comments. To further improve the performance of DeepCS, Huang et al. \cite{huang2020code} incorporated code/query embedding with an attention mechanism. Shuai et al. \cite{shuai2020improving} leveraged a co-attention mechanism to learn the correlation between the embedded query and code. Wan et al. \cite{wan2019multi} proposed a tool with more code representations, which embeds the code structure by a tree-based LSTM and the call graph of code by a GGNN (Gate-Graph Neural Network). Recently, researchers \cite{yao2019coacor,ye2020leveraging} improved code search with generated code summarization, and then built tools as a reinforcement learning process of code search and code summarization tasks. The generated summarization is important because it is an abstraction of the code and shares the same semantic level as a developer's search query.

However, the above tools are complex, and training them is time-consuming. Therefore, researchers also investigated more lightweight tool approaches at the same time. Sachdev et al. \cite{sachdev2018retrieval} proposed a tool, NCS, that leveraged an unsupervised token-level embedding fastText \cite{bojanowski2017enriching} to transform code and query into vectors. NCS searched candidate code for a query by using the TF-IDF weighting method \cite{haiduc2013automatic}, and finally re-ranked the candidates by comparing their cosine similarity to the query vector. To improve the search effectiveness of NCS, Liu et al. \cite{liu2019neural} added a query expansion technique to the NCS; Sachdev et al. \cite{sachdev2018retrieval} replaced the unsupervised component TF-IDF in NCS by a neural network with an attention mechanism. This can be trained by the code-comment pairs as DeepCS \cite{gu2018deep}. 

\subsubsection{Code Clone Search}\label{code_clone_search}

A code clone search tool takes a piece of code as a query and returns a list of similar codes \cite{ragkhitwetsagul2019siamese,luan2019aroma,kim2018facoy}. It differs from code clone detection \cite{roy2007survey,sheneamer2016survey} because it is query-centric, retrieves only clones that are associate with the query, instead of looking for a complete set of clone pairs in the codebase as the clone detection, and cares about the tool scalability \cite{ragkhitwetsagul2019siamese}. Table \ref{tab_ui_code_search} shows the reviewed studies related to the code clone search task. 

Early tools regarded code clone search as a token-by-token matching issue. CCFinder \cite{kamiya2002ccfinder} identifies whether the partial token sequences of a candidate code snippet contains the target query code. SourcererCC \cite{sajnani2016sourcerercc} calculated code similarity based on the overlap degree between the tokens of two codes. When the degree value is lower than a pre-defined coefficient, SourcererCC returns the code in a codebase as a clone. However, code does not just consist of tokens but with particular structures. Thus, further tools considered this code structural information. Lee et al. \cite{lee2010instant} transformed a source code snippet into a control flow graph (CFG). The graph is represented by a characteristic vector via the locality sensitive hashing (LSH) algorithm \cite{gionis1999similarity,jiang2007deckard}, where a node in CFG is composed by its subgraphs. Code clone search is then regarded as a subgraph matching problem. Balachandran \cite{balachandran2015query} improved the LSH representation method by incorporating the more code structural feature, namely the abstract syntax trees (ASTs). 

In recent years, researchers have built tools based on learning models. DLC \cite{white2016deep} leveraged a deep learning technique to embed binary code into vectors and learned their lexical relationship. Code clone search can then be performed by measuring the cosine similarity between code vectors. Additionally, to learn the higher semantic difference between code, TBCAA \cite{chen2019capturing} applies the tree-based convolution neural network to capture the structural feature, namely the AST (abstract syntax tree) of code. Researchers have also sought other ways to improve code clone search. Siamese \cite{ragkhitwetsagul2019siamese} incorporates a multi-representation, corresponding to four clone types, to represent indexed corpus of code, improves the query quality by leveraging the knowledge of token frequency in the codebase, and finally re-ranks the searched candidate code based on the TF-IDF weighting method. FaCoY \cite{kim2018facoy} extended the query with related code in Stack Overflow, and searches similar code fragments against the code index built from the source code of software projects. Aroma \cite{luan2019aroma} pruned and clustered candidate code, and intersects the snippets in each cluster to carve out a maximal code snippet. This snippet is common to all the snippets in the cluster and which contains the query snippet. The set of intersected code snippets are then returned as recommended code snippets.

\begin{table}[h]
    \centering
    \footnotesize
    \caption{Code clone search tools.}
    \setlength{\tabcolsep}{10pt}{
    \begin{tabular}{|cccc|}
        \toprule
        \textbf{No.} & \textbf{Year} & \textbf{Tool Name} & \textbf{Baselines} \\
        \midrule
        1&2019 & TBCAA \cite{chen2019capturing} & Siamese, SourcererCC, CCFinder, DLC\\
        2&2019 & Siamese \cite{ragkhitwetsagul2019siamese} & FaCoy, SourcererCC, CCFinder\\
        3&2019 & Aroma \cite{luan2019aroma} & SourcererCC\\
        4&2018 & FaCoY \cite{kim2018facoy} & SourcererCC, CCFinder\\
        5&2016 & DLC \cite{white2016deep} & -\\
        6&2016 & SourcererCC \cite{sajnani2016sourcerercc} & CCFinder\\
        7&2015 & Balachandran15 \cite{balachandran2015query} &  -\\
        8&2010 & Lee10 \cite{lee2010instant} & -\\
        9&2002 & CCFinder \cite{kamiya2002ccfinder} &  -\\
        \bottomrule
    \end{tabular}}
    \label{tab_ui_code_search}
\end{table}

\subsubsection{I/O Example Code Search}

For I/O example code search, a query contains a set of examples specifying the desired input/output (I/O) behaviors of target code \cite{chen2020enhancing}. The given I/O examples reflect incomplete functional specifications that can be collected from development requirements \cite{chen2020enhancing} or test cases \cite{lemos2011test}. An I/O example code search tool aims to find the code methods that match a specified I/O example. How the code behaves is unimportant \cite{chen2020enhancing}. Table \ref{tab_example_based_code_search} shows I/O example code search tools included in this systematic review. We can see that these tools have mainly refined the results of existing online search engines, such as GitHub search, Google Code, and Sourcerer.

To find a method that matches the expected I/O examples, early tools \cite{sahavechaphan2006xsnippet,thummalapenta2007parseweb,reiss2009semantics} employed graph-based code mining algorithms to mine paths that start with the input example and end with the output example. However, it was observed although some methods did not satisfy the I/O requirement, their partial code meets the expectation. Therefore, researchers \cite{sun2019slicing,lemos2011test} leveraged slicing techniques that locate the output example from a method and extract related code snippets backwards. Such a technique excludes the methods that cannot trace the input example. To improve search performance, researchers also leveraged query processing techniques to optimize the initial results of online search engines, namely expanding query with appropriate synonyms \cite{lemos2014thesaurusExpand,lemos2014thesaurus}.

Dynamic analysis techniques have also been adopted for I/O example code search. Stolee et al. \cite{stolee2016code} proposed a tool called Satsy. It is based on a symbolic execution approach and works in two phases. During an offline encoding phase, Satsy encodes the semantics of code in codebase into logical formulas concerning their input/output variables. During an online search, Satsy binds concrete values from I/O examples to compatible variables in each formula to construct a constraint, and checks the satisfiability of the constraint using a solver Z3 \cite{de2008z3}. Although Satsy has been applied to search for Java code in previous studies \cite{stolee2016code}, its usefulness in daily code search activities is limited, as it handles only loop-free code snippets manipulating data of char, int, boolean, and String types. To extend the usefulness of Satsy, Chen et al. \cite{chen2020enhancing} proposed a tool Quebio. Different from Satsy, its symbolic encoding phase supports more language features like the invocation to library APIs, which enables Quebio to handle more data types (e.g., array, List, Set, and Map) during the search. This new feature enables Quebio to be used in a wider range of scenarios. 

\begin{table}[h]
    \centering
    \footnotesize
    \caption{I/O example code search tools.}
    \setlength{\tabcolsep}{20pt}{
    \begin{tabular}{|cccc|}
        \toprule
        \textbf{No.} & \textbf{Year} & \textbf{Tool Name} &  \textbf{Baselines} \\
        \midrule
        1&2020 & Quebio \cite{chen2020enhancing} & Satsy \\
        2&2019 & TIRSnippet \cite{sun2019slicing} & PARSEWeb \\
        3&2016 & Satsy \cite{stolee2016code} & - \\
        4&2014 & Lemos14 \cite{lemos2014thesaurus} & QE$_{wct}$ \\
        5&2014 & QE$_{wct}$ \cite{lemos2014thesaurusExpand} & CodeGenie \\
        6&2011 & CodeGenie \cite{lemos2011test} & - \\
        7&2009 & S6 \cite{reiss2009semantics} & -\\
        8&2007 & PARSEWeb \cite{thummalapenta2007parseweb} & - \\
        \bottomrule
    \end{tabular}}
    \label{tab_example_based_code_search}
\end{table}

\subsubsection{API-Based Code Search}

Developers write code using various APIs, but only a limited portion is explained with code examples, where only around 2\% of APIs in JDK 5 (27k in total) provide examples \cite{zhong2009mapo,wang2013mining}. Therefore, developers have to type the expected API in existing search engines, such as Google. However, the search engine often returns numerous results and most of them do not meet developers' expectations \cite{zhang2019enriching,gu2019codekernel}. Thus, many tools have been proposed to mitigate developers' inspection efforts by clustering and ranking candidate code \cite{kim2010towards,mar2011recommending,wang2013mining}. 

One early example is MAPO \cite{zhong2009mapo} that searches a list of code relevant to the target API and clusters the code according to their API call sequences using a classical hierarchical clustering technique \cite{han2011data}. To help developers find expected code quickly, MAPO also generates code call patterns (i.e., a sequence of API calls) for describing each cluster. As the major challenge is how to cluster and rank the searched code, researchers have proposed a number of solutions. EXoaDocs \cite{kim2010towards} transforms the searched code into a vector space according to their AST structure, clusters them by using a hierarchical clustering algorithm (the centroid of a cluster is regarded as the representative code), and ranks the code based on three factors. These are representativeness -- the reciprocal of the similarity to the representative code of the corresponding cluster; conciseness -- the reciprocal of code length; and correctness -- the degree the code is related to the target API. PropER-Doc \cite{mar2011recommending} clusters candidate code based on their interacted API types, and ranks the candidates based on three designed metrics: significance, how the API in code related to the query; density, the portion of code lines that refers to the query; cohesiveness, the aggregation level of the query described within the code. UPMiner \cite{wang2013mining} clusters code based on the similarity of the API sequences and groups the frequent closed sequences into a code pattern for a cluster using the tool BIDE \cite{wang2004bide}. The code pattern that covers more possible APIs and contains fewer redundant lines is ranked higher. MUSE \cite{moreno2015how} discards irrelevant lines of code in the codebase, clusters the simplified code by a code clone detection method \cite{harris2003simian}, and ranks the representative code in clusters based on their reusability, understandability, and popularity.

ADECK \cite{zhang2019enriching} collects candidate code examples from the community question and answer (Q\&A) forum Stack Overflow. It represents the searched post as a tuple consisted of the post title and the best-answered code scored by users. Then ADECK clusters the tuples based their semantic similarities by leveraging the APCluster method \cite{frey2007clustering}. KodeKernel \cite{gu2019codekernel} represents a source code as an object usage graph, instead of method invocation sequences or feature vectors. It clusters code by embedding them into a continuous space using a graph kernel.  KodeKernel then selects a representative code from each cluster based on two designed ranking metrics: centrality, the average distance from one code to another in the cluster; and specificity, the code contains less rarely appeared lines.

\begin{table}[h]
    \centering
    \footnotesize
    \caption{API-based code search tools.}
    \setlength{\tabcolsep}{20pt}{
    \begin{tabular}{|cccc|}
        \toprule
        \textbf{No.} & \textbf{Year} & \textbf{Tool Name} & \textbf{Baselines} \\
        \midrule
        1&2019 & KodeKernel \cite{gu2019codekernel} & eXoaDocs, MUSE\\
        2&2019 & ADECK \cite{zhang2019enriching} & eXoaDocs\\
        3&2015 & MUSE \cite{moreno2015how}& -\\
        4&2013 & UPMiner \cite{wang2013mining} & MAPO \\
        5&2011 & PropER-Doc \cite{mar2011recommending}& - \\
        6&2010 & eXoaDocs \cite{kim2010towards} & -\\
        7&2009 & MAPO \cite{zhong2009mapo} & -\\
        \bottomrule
    \end{tabular}}
    \label{tab_example_based_code_search}
\end{table}

\subsubsection{Binary Code Search}

When deploying source code on different operating systems with various compilers and optimization methods, source code is usually transformed into different binary codes. For a given binary code, how to search for the same binary code but compiled in other forms becomes the objective of the binary code search task. Assembly code (i.e., the human-readable binary code) is commonly used to build the codebase \cite{ding2019asm2vec,khoo2013rendezvous}. This is important for plagiarism detection, malware detection, and software vulnerability auditing \cite{khoo2013rendezvous,david2014tracelet}. 

To address the binary code search task, Rendezvous \cite{khoo2013rendezvous} compares the descriptive statistics between code tokens in terms of the mnemonic n-grams, mnemonic n-perms, control flow sub-graph, and data constants. To ensure tool scalability, Rendezvous builds indexing for a codebase to reduce the scope of search space according to the given query. However, the low-level compiler transformations can strongly affect the performance of Rendezvous. To tackle this issue, Tracy \cite{david2014tracelet} decomposes code into tracelets -- the continuous, short, partial traces of an execution -- and compares these tracelets based on a Jaccard containment similarity \cite{agrawal2010indexing} in the face of low-level compiler transformations. Kam1n0 \cite{ding2016kam1n0} represents the control flow graph (CFG) of binary code as a LSH (locality sensitive hashing) scheme \cite{kulis2009kernelized}, where a node in CFG is formed as a combination of its subgraph. Binary code search is then transformed into a subgraph search problem. It is challenging to align the semantics between two binary codes. To overcome this challenge, BingGo-E \cite{xue2018accurate} selectively inlines a binary code with relevant libraries and user-defined codes to complete the semantics in code. Asm2Vec \cite{ding2019asm2vec} leverages a deep learning technique to jointly learn the semantic relationships between binary code. The learned code representation can largely mitigate the manual incorporation of the complex prior domain knowledge.

\begin{table}[h]
    \centering
    \footnotesize
    \caption{Binary clone search tools.}
    \setlength{\tabcolsep}{20pt}{
    \begin{tabular}{|cccc|}
        \toprule
        \textbf{No.} & \textbf{Year} & \textbf{Tool Name}   & \textbf{Baselines} \\
        \midrule
        1&2019 & Asm2Vec \cite{ding2019asm2vec} & Rendezvous \\
        2&2018 & BingGo-E \cite{xue2018accurate}& Tracy \\
        3&2016 & Kam1n0 \cite{ding2016kam1n0} & Rendezvous, Tracy \\
        4&2014 & Tracy \cite{david2014tracelet} & Rendezvous \\
        5&2013 & Rendezvous \cite{khoo2013rendezvous} & -\\
        \bottomrule
    \end{tabular}}
    \label{tab_binary_code_search}
\end{table}

\subsubsection{UI Code Search}

When developing software user interfaces (UIs) developers commonly draft UI sketches and implement corresponding UI code with related APIs. This often takes enormous efforts. UI code search tools take UI sketches as a query and search for UI code snippets that match the requirement of the UI sketch from a codebase. Table \ref{tab_ui_code_search} shows three UI code search tools we reviewed in this study. Reiss et al. \cite{reiss2018seeking} converted the image of a developer's UI sketch into a scalable vector graphic (SVG) diagram, and reduced the search space by using an existing text-based search engine $S^{6}$ \cite{jiang2018semantics} with related keywords. After a series of transformations for the candidate code, this tool returns the ones that can be compiled and run. Behrang et al. \cite{behrang2018guifetch} proposed a similar tool GUIFetch but measures the query-code relevancy with a comprehensive metric based on the screen similarity in terms of the screen type, size, and position, and the screen transition similarity. To improve query representation, Xie et al. \cite{xie2019user} adapted the pix2code \cite{beltramelli2018pix2code} tool to automatically extract the code structure from a UI sketch. Pix2code uses a DL model that can capture the UI components types and their hierarchical relationships, which was trained with manually labeled UI sketches. To sort candidate UI code, Xie et al. \cite{xie2019user} measured the layout distance between query and candidate code based on the Levenshtein distance \cite{levenshtein1966binary}. 

\begin{table}[h]
    \centering
    \footnotesize
    \caption{UI code search tools.}
    \setlength{\tabcolsep}{20pt}{
    \begin{tabular}{|cccc|}
        \toprule
        \textbf{No.} & \textbf{Year} & \textbf{Tool Name} &  \textbf{Baselines} \\
        \midrule
        1&2019 & Xie19 \cite{xie2019user} & Reiss18 \\
        2&2018 & GUIFetch \cite{behrang2018guifetch} & -\\
        3&2018 & Reiss18 \cite{reiss2018seeking}& -\\
        \bottomrule
    \end{tabular}} 
    \label{tab_ui_code_search}
\end{table}

\subsubsection{Programming Video Search}

Programming video code snippet search is a new code search task recently investigated by Bao et al. \cite{bao2020psc2code}. This task aims to search for relevant code snippets in programming videos (e.g., from YouTube) using text-based queries. The major challenge is how to capture the code in videos and transform it into text. Code search can then be implemented by using text-based code search tools described earlier in Section \ref{text_based_code_search}. To capture the relevant code frame in programming videos, Bao et al. \cite{bao2020psc2code} proposed a tool psc2code. It removes noisy frames by a CNN (Convolutional Neural Network) based image classification, extracts source code by calling a professional ORC (Optical Character Recognition) tool, and performs code search by using the TF-IDF algorithm.

\subsection{Replication Packages}\label{replication}

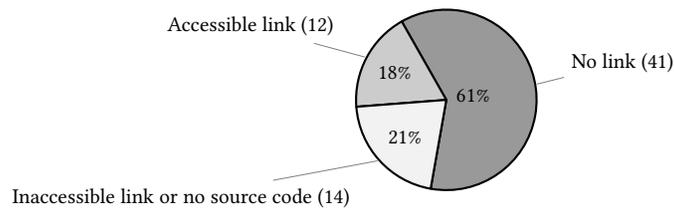
\begin{figure}[h]
    \centering
    \footnotesize
    \begin{tikzpicture}[scale=0.4]
    \pie[rotate=260, text=pin, color={gray!80, gray!40, gray!10}]{61/No link (41), 18/Accessible link (12), 21/Inaccessible link or no source code (14)}
    \end{tikzpicture}
    \caption{Replication package.}
    \label{fig_replicability}
\end{figure}

We wondered how often the reviewed 67 code search tools share replication packages in their papers. We searched and inspected all the links in each paper. If a replication package link is available, we checked the link accessibility and whether the replication package contains source code. If we found no relevant link in a paper, we also searched its replication package in GitHub with the paper title. The pie chart in Fig. \ref{fig_replicability} shows that only 18\% of the reviewed studies provide accessible replication packages. Among the other studies, 14 studies provide inaccessible links in the paper or accessible links without source code. We found no description of replication packages for 41 studies. To facilitate future code search study, we provide the usable replication packages in Table \ref{tab_urls}.

\begin{table}[h]
    \centering
    \footnotesize
    \caption{Replication package links.}
    \setlength{\tabcolsep}{19.5pt}{
    \begin{tabular}{|lll|}
    \toprule
        \textbf{Task} & \textbf{Tool} & \textbf{URL} \\
    \midrule
         \multirow{4}{*}{Text-Based Code Search}
         & CARLCS-CNN & https://github.com/cqu-isse/CARLCS-CNN \\
         & CoaCor & https://github.com/LittleYUYU/CoaCor\\
         & DeepCS & https://github.com/guxd/deep-code-search\\
         & CODE-NN & https://github.com/sriniiyer/codenn\\
         \midrule
         I/O example Code Search & MUSE & https://github.com/lmorenoc/icse15-muse-appendix\\
         \midrule
         API-Based Code Search & - & - \\
         \midrule
         \multirow{3}{*}{Code Clone Search} 
         & Siamese & https://github.com/UCL-CREST/Siamese\\
         & Aroma & https://github.com/facebookresearch/aroma-paper-artifacts\\
         & FaCoY & https://github.com/FalconLK/facoy\\
         \midrule
         \multirow{2}{*}{Binary Clone Search}
         & Kam1n0 & https://github.com/McGill-DMaS/Kam1n0-Community\\
         & Tracy & https://github.com/Yanivmd/TRACY\\
         \midrule
         UI Search & Xie19 & https://github.com/yingtao-xie/code\_retrieval/\\
         \midrule
         Programming Video Search & psc2code & https://github.com/baolingfeng/psc2code \\
    \bottomrule
    \end{tabular}}
    \label{tab_urls}
\end{table}






\begin{framed}
\filbreak
\noindent\textit{\textbf{Summary of answers to RQ2:}}
\begin{itemize}
\item \textit{74\% of tools searched source code with queries written in natural language (i.e., text, API name, input/output example).}\vspace{3pt}
\item\textit{Deep learning is the most popular modeling technique in the last two years.}\vspace{3pt}
\item\textit{Inverted Index was frequently used for accelerating code search efficiency; researchers also leveraged other auxiliary techniques (i.e., query reformulation, code clustering, active learning) to improve the search accuracy.}\vspace{3pt}
\item\textit{We found only 12 code search studies shared accessible replication package links in their papers or provided source code in GitHub.}
\end{itemize}
\end{framed}

\section{RQ3: How do studies evaluate code search tools?}\label{rq3}
We investigated the key approaches used to evaluate the 67 code search tools including aspects of the codebase, query, evaluation method, and performance measures.    

\subsection{Codebase}\label{codebase}

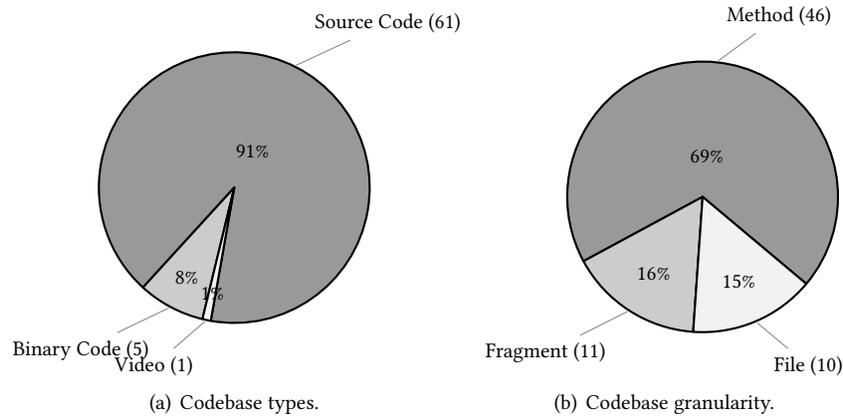
\begin{figure}[h]
    \centering
    \footnotesize
    \subfigure[Codebase types.]{
    \begin{tikzpicture}[scale=0.6]
    \pie[rotate=260, text=pin, color={gray!80, gray!40, gray!10}]{91/Source Code (61), 8/Binary Code (5), 1/Video (1)}
    \end{tikzpicture}}
    \subfigure[Codebase granularity.]{
    \begin{tikzpicture}[scale=0.6]
    \pie[rotate=320, text=pin, color={gray!80, gray!40, gray!10}]{69/Method (46), 16/Fragment (11), 15/File (10)}
    \end{tikzpicture}}
    \caption{Codebase types and granularity.}
    \label{fig_codebase_type_granularity}
\end{figure}

Various characteristics of the codebase define the search space of a specific code search type. After inspecting the 67 code search tools, we found three types of codebases. As illustrated in Fig. \ref{fig_codebase_type_granularity}(a), 91\% of codebases are built with source code written by developers in high-level programming languages (e.g., Java); 8\% of codebases consist of binary code, i.e., the compiled source code; and one study collected a corpus of programming tutorial video with source code as a codebase to help developers search code from videos \cite{bao2020psc2code}. Fig. \ref{fig_codebase_type_granularity}(b) shows the distribution of the code granularity: 69\% of the tools regard methods as search targets; 16\% of the tools focus on recovering code fragments; and 15\% of the tools concentrate on retrieving relevant files, e.g. the app UI code \cite{reiss2018seeking}. 

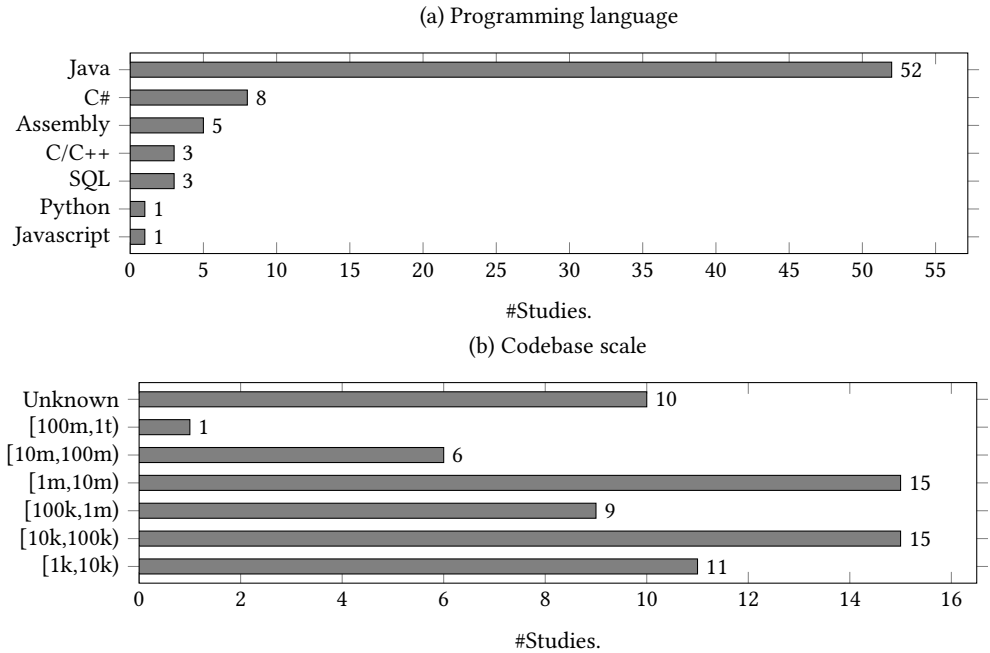
\begin{figure}[h]
    \centering
    \small
    \begin{tikzpicture}
    \begin{axis}[
        xbar, 
        label style={font=\small}, 
        tick label style={font=\small},
        width = 0.8\linewidth,
        height=4.25cm,
        bar width=0.2cm,
        xmin=0,
        xlabel={\#Studies.},
        title={(a) Programming language},
        symbolic y coords={{Javascript},{Python},{SQL},{C/C++},{Assembly},{C\#},{Java}}, 
        ytick=data,
        nodes near coords, 
        nodes near coords align={horizontal},]
        \addplot[black, fill=gray] coordinates {(1,{Javascript})(1,{Python})(3,{SQL})(3,{C/C++})(5,{Assembly})(8,{C\#})(52,{Java})};
    \end{axis}
    \end{tikzpicture}
    \begin{tikzpicture}
    \begin{axis}[
        xbar, 
        label style={font=\small}, 
        tick label style={font=\small},
        width = 0.8\linewidth,
        height=4.25cm,
        bar width=0.2cm,
        xmin=0,
        xlabel={\#Studies.},
        title={(b) Codebase scale},
        symbolic y coords={{[1k,10k)},{[10k,100k)},{[100k,1m)},{[1m,10m)},{[10m,100m)},{[100m,1t)},{Unknown}}, 
        ytick=data,
        nodes near coords, 
        nodes near coords align={horizontal},]
        \addplot[black, fill=gray] coordinates {(11,{[1k,10k)})(15,{[10k,100k)})(9,{[100k,1m)})(15,{[1m,10m)})(6,{[10m,100m)})(1,{[100m,1t)})(10,{Unknown})};
    \end{axis}
    \end{tikzpicture}
    \caption{Codebase language and scale.}
    \label{fig_codebase_language_scale}
\end{figure}

Fig. \ref{fig_codebase_language_scale} (a) shows the distribution of programming language in code bases: Java is the most favored language with 52 studies, followed by C\#, Assembly, C/C++, SQL, Python, and Javascript. Fig. \ref{fig_codebase_language_scale}(b) shows the distribution of the codebase scale in terms of the number of studied instances (i.e., method, fragment, or file). We can see that 35 of the selected studies used small scale codebases with more than 1k and lower than 1m code items. 22 studies constructed larger scale codebases with more than one million code items. However, the codebase scales of ten studies are unknown because their tools are based on online search engines, such as the GitHub search\footnote{https://github.com/search} with millions of open source repositories. Table \ref{tab_scale_granularity} shows the codebase scale in different code granularities (method, fragment, and file) respectively. We can see that most studies that search code methods have built larger codebases ($>$10m).


Table \ref{tab_codebase_source} summarizes the data sources of different codebases. By classifying the sources into four categories, we can see that the open source community is the biggest source category including GitHub \cite{gu2018deep}, SourceForge \cite{zhong2009mapo}, Google code \cite{reiss2009semantics}, Apache \cite{lee2018comment}, FDroid \cite{jiang2016rosf}, OpenHub \cite{li2016relationship}, and Tigris.org \cite{lee2010instant}. Among these communities, GitHub is the most popular one related to 23 code search studies. Researchers also collected data from app stores (Google Play and Apple store \cite{xie2019user}), enterprise with closed projects (Microsoft \cite{wang2013mining} and Amazon \cite{li2016relationship}), and programming forum and videos (Stack Overflow \cite{allamanis2015bimodal,zhang2019enriching} and YouTube \cite{bao2020psc2code}). 19\% of studies manually selected and downloaded some projects according to their experience \cite{mcmillan2013portfolio,bajracharya2010leveraging}, while 13\% of studies proposed new techniques to better support online search engines (e.g., GitHub search \cite{gu2019codekernel} and Google code \cite{mar2011recommending}). 

\begin{table}[h]
    \centering
    \footnotesize
    \caption{Codebase scale in different code granularities (i.e., method, fragment, and file).}
    \begin{tabular}{|c|cccccccc|}
        \toprule
        \textbf{Scale} & \textbf{Unknown} & \textbf{$[$100m,1t$)$} & \textbf{$[$10m,100m$)$} & \textbf{$[$1m,10m$)$} & \textbf{$[$100k,1m$)$} & \textbf{$[$10k,100k$)$} & \textbf{$[$1k,10k$)$} & \textbf{Total} \\
        \midrule
        Method & 7 & 1 & 6 & 7 & 6 & 12 & 7 & 46\\
        Fragment & 0 & 0 & 0 & 6 & 2 & 1 & 2 & 11\\
        File & 3 & 0 & 0 & 2 & 1 & 2 & 2 & 10\\
        \textbf{Total} & 10 & 1 & 6 & 15 & 9 & 15 & 11 & 67\\
        \bottomrule
    \end{tabular}
    \label{tab_scale_granularity}
\end{table}

\begin{table}
    \centering
    \footnotesize
    \caption{Codebase source.}
    \setlength{\tabcolsep}{23.8pt}{
    \begin{tabular}{|llcc|}
        \toprule
        \textbf{Source Type} & \textbf{Source} & \textbf{\#Studies} & \textbf{Percent}\\
        \midrule
        \multirow{7}{*}{Public Code Repositories} & GitHub & 23&29\%\\
         & SourceForge & 6&8\%\\
         & Google Code & 4&5\%\\
         & Apache & 3&4\%\\
         & FDroid & 2&3\%\\
         & OpenHub & 1&1\%\\
         & Tigris.org & 1&1\%\\
        \midrule
        \multirow{2}{*}{App Store} & Google Play & 1&1\%\\
         & Apple Store & 1&1\%\\
        \midrule
        \multirow{2}{*}{Internal Code Repositories} & Microsoft & 1&1\%\\
         & Amazon & 1&1\%\\
        \midrule
        \multirow{2}{*}{Forum and Video Sharing Platform} & Stack Overflow & 8&10\%\\
        &YouTube & 1&1\%\\
        \midrule
        \multirow{2}{*}{Other}&Selected Projects & 15 &19\%\\
        &Online Search Engine& 10&13\%\\
        \bottomrule
    \end{tabular}}
    \label{tab_codebase_source}
\end{table}

\subsection{Queries}\label{query}

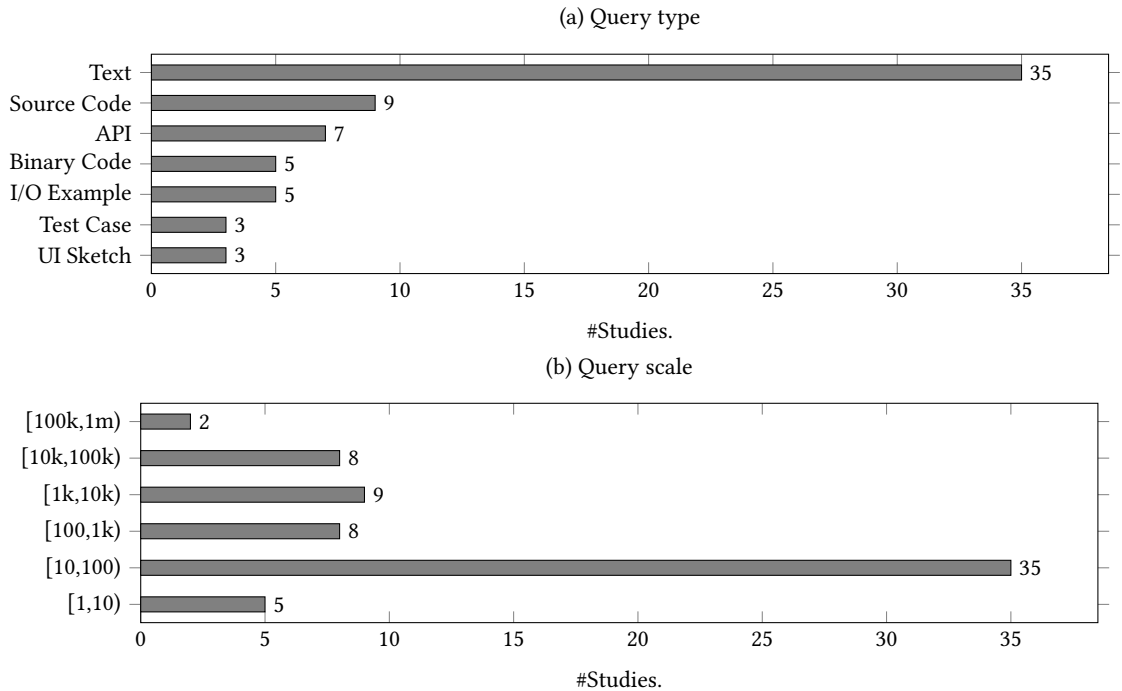
\begin{figure}[h]
    \centering
    \small
    \begin{tikzpicture}
    \begin{axis}[
        xbar, 
        label style={font=\small}, 
        tick label style={font=\small},
        width = 0.9\linewidth,
        height=4.5cm,
        bar width=0.2cm,
        xmin=0,
        xlabel={\#Studies.},
        title={(a) Query type},
        symbolic y coords={{UI Sketch},{Test Case},{I/O Example},{Binary Code},{API},{Source Code},{Text}}, 
        ytick=data,
        nodes near coords, 
        nodes near coords align={horizontal},]
        \addplot[black, fill=gray] coordinates {(3,{UI Sketch})(3,{Test Case})(5,{I/O Example})(5,{Binary Code})(7,{API})(9,{Source Code})(35,{Text})};
    \end{axis}
    \end{tikzpicture}
    \begin{tikzpicture}
    \begin{axis}[
        xbar, 
        label style={font=\small}, 
        tick label style={font=\small},
        width = 0.9\linewidth,
        height=4.5cm,
        bar width=0.2cm,
        xmin=0,
        xlabel={\#Studies.},
        title={(b) Query scale},
        symbolic y coords={{[1,10)},{[10,100)},{[100,1k)},{[1k,10k)},{[10k,100k)},{[100k,1m)}}, 
        ytick=data,
        nodes near coords, 
        nodes near coords align={horizontal},]
        \addplot[black, fill=gray] coordinates {(5,{[1,10)})(35,{[10,100)})(8,{[100,1k)})(9,{[1k,10k)})(8,{[10k,100k)})(2,{[100k,1m)})};
    \end{axis}
    \end{tikzpicture}
    \caption{Query type and scale.}
    \label{fig_query_type_scale}
\end{figure}

Code search queries reflect developers' search requirements, and their features determine how a code search tool can support developers' intents. Fig. \ref{fig_query_type_scale}(a) shows that there are seven types of queries: \textit{1) text} -- a short string of text written in natural language, e.g., "how to convert inputstream to a string"; \textit{2) source code} -- a code method or fragment to retrieve similar code from codebase; \textit{3) API} -- an official or third-party API name to search the API usage example; \textit{4) binary code } -- an assembly code (i.e., the compiled source code) with similar search task as the query of source code; \textit{5) I/O Example} -- the input and output variable types or examples for a code method; \textit{6) Test case} -- a piece of testing code; and \textit{7) UI sketch} -- an image of UI skeleton drafted by UI designers. Among these query choices, the text query can be supported by 35 code search studies. The query-based search is popular because it is be regarded as a general search engine like the GitHub search. 

Fig. \ref{fig_query_type_scale}(b) illustrates the distribution of query scales in the reviewed 67 code search tools. We can see that 35 studies evaluated tool effectiveness with 10-100 queries. Five studies tested tools with no more than ten queries, while 25 studies performed code search with larger scale of inputs of  more than 100 queries. 

We also analyzed the query sources adopted by different code search studies. As illustrated in Table \ref{tab_query_source}, 39\% of queries were collected from question and answer forums including Stack Overflow \cite{cambronero2019deep} and Eclipse FAQs \cite{bajracharya2010leveraging}; 6\% of queries were extracted from software development kits, including JDK \cite{gu2019codekernel}, Android development kit \cite{zhang2019enriching}, and .Net framework \cite{wang2013mining}. The remaining 55\% of queries were manually selected from frequently used examples: codebase \cite{thummalapenta2007parseweb,lemos2011test}, search logs of practically used code search engine \cite{martie2015codeexchange,yang2017iecs}, a summary of search on the Internet \cite{mcmillan2013portfolio,reiss2018seeking}, and automatically generated queries \cite{zou2020query,huang2018query}.

\begin{table}[h]
    \centering
    \footnotesize
    \caption{Query source.}
    \setlength{\tabcolsep}{23pt}{
    \begin{tabular}{|llcc|}
        \toprule
        \textbf{Type} & \textbf{Source} & \textbf{\#Studies} & \textbf{Percent}\\
        \midrule
        \multirow{2}{*}{Question and Answer Forum} 
        & Stack Overflow & 25 & 36\%\\
        & Eclipse FAQs & 2 & 3\%\\
        \midrule
        \multirow{3}{*}{Development Kit} 
        & Java Development Kit & 3 & 4\%\\
        & Android Development Kit & 1 & 1\%\\
        & .NET Framework & 1 & 1\%\\
        \midrule
        \multirow{4}{*}{Frequently Used Examples}
        & Codebase & 30 & 43\%\\
        & Search Log & 4 & 6\%\\
        & Web Search & 2 & 3\%\\
        & Automatic Generation & 2 & 3\%\\
        \bottomrule
    \end{tabular}}
    \label{tab_query_source}
\end{table}

\subsection{Evaluation Methods}\label{evaluation}

The reviewed code search studies have evaluated new tools in four ways, as listed in Table \ref{tab_evaluation_method}. Manual identification is the major evaluation method in 37 related studies. In this case, researchers or invited developers manually inspected the top-$n$ result list returned by a code search tool and identified their relevancy to query intent one by one. However, such manual evaluation approaches may suffer from subjective bias. To mitigate this issue, 30 code search studies provided three approaches to automatically identify the relevancy between queries and searched results. 19 studies constructed a ground-truth where selected code snippets correspond to each query result in advance. However, obtaining the ground-truth is not easy and is very manually intensive to do. Therefore, two studies \cite{gu2019codekernel,xie2019user} asked researchers or developers to manually annotate the ground-truth in the codebase, which requires the codebase scale to be amenable to these limited manual efforts. To mitigate issues when using manual efforts, nine studies designed a measurement to score the query-code relevancy (e.g., leveraging a clone detection method to score the similarity between a search code and an example code \cite{huang2019deep,sachdev2018retrieval}) and determined relevancy if the score is larger than a pre-defined threshold. However, choosing the right threshold is difficult and researchers have tried their best to simulate manual identification. 

\begin{table}[h]
    \centering
    \footnotesize
    \caption{Evaluation method.}
    \setlength{\tabcolsep}{10pt}{
    \begin{tabular}{|clL{0.52\linewidth}rr|}
    \toprule
        \textbf{No.} &\textbf{Method} & \textbf{Description} & \textbf{Count} & \textbf{Percent}\\
    \midrule
        1&Manual & Manual identification &37&55\%  \\
        2&Automated1 & Automated identification based on collected query-code pairs &19&28\%\\
        3&Automated2 & Automated identification based on manually annotated query-code relevancy &2&3\%\\
        4&Automated3 & Automated identification based on a designed algorithm to judge the query-code relevancy &9&14\%\\
        -&\textbf{Total} & - & 67&100\%\\
    \bottomrule
    \end{tabular}}
    \label{tab_evaluation_method}
\end{table}

\subsection{Performance Metrics}\label{performance}

\begin{table}
    \centering
    \footnotesize
    \caption{Performance Metrics.}
    \setlength{\tabcolsep}{5pt}{
    \begin{tabular}{|l|lL{0.6\linewidth}c|}
    \toprule
        \textbf{Type} & \textbf{Metric} & \textbf{Description} & \textbf{\#Studies}  \\
    \midrule
        \multirow{21}{*}{Ranking} 
        &Precision@k& The percentage of relevant code in the top-k result list for each query. &27\\
        &MRR@k& The average of reciprocal ranks of the results of a set of queries, where the reciprocal rank of a query is the inverse rank of first relevant code in the top-k result list. &19\\
        &NDCG@k& The normalized discounted cumulative gain for the top-k search results. &16\\
        &SuccessRate@k& The proportion of queries that the relevant code could be found in the top-k result lists. &11\\
        &MAP@k& Mean average precision, where the average precision of a query is the mean of precision at each rank. &5\\
        &FRank@k & The rank of the first relevant code in the top-k result list for a given query. &4\\
        &FFP@k & The rank of the first false positive code in the top-k results for a given query. &1\\
        &MSE@k & The mean squared errors between the relevancy ratings of each search code and the ground-truth ratings. &1\\
        &KCC@k & The Kendall's correlation coefficient between the searched code rankings and the ground-truth rankings. &1\\
        &SCC@k & The Spearman's correlation coefficient between the searched code rankings and the ground-truth rankings. &1\\
        &PCC@k & The Pearson correlation coefficient between the searched code rankings and the ground-truth rankings.&1\\
        \midrule
        \multirow{8}{*}{Classification}
        &Precision & The proportion of code in a result list are truly relevant for a query. & 16 \\
        &Recall & The proportion of relevant code in codebase that are correctly retrieved for a query. & 9 \\
        &F1-score & A harmonic mean of Precision ($p$) and Recall ($r$), equaling to $2pr/(p+r)$. & 4\\
        &AUROC & The area under the receiver operating characteristics curve. &2\\
        &F2-score & A weighted harmonic mean of Precision ($p$) and Recall ($r$) with double weights on recall, equaling to $5pr/(4p+r)$. &1\\
        &AUPR & The area under the precision-recall curve. &1\\
        \midrule
        Other
        &Search Time & The average time duration for each code search query to be processed. &13\\
    \bottomrule
    \end{tabular}}
    \label{tab_performance}
\end{table}

To measure the effectiveness of a code search task, studies used various performance metrics as listed in Table \ref{tab_performance}. Generally, there are two types of accuracy measure approaches -- one treating code search as a ranking issue and the other treating code search as a  classification issue.

For ranking, for a search result list with top-$k$ code, a metric measures the accuracy concerning the important location in the list \cite{gu2018deep,shuai2020improving,liu2020simplifying}. FRank@k and FFP@k are metrics for an individual query, which are the ranks of the first relevant code and the first false positive code respectively. To measure the comprehensive search accuracy for all queries, MRR@k and SuccessRate@k care about the first correctly search code for a query, where MRR@k is an average of reciprocal ranks of the first correct search for each query; SuccessRate@k counts the percentage of queries that retrieved at least one relevant code. To measure the accuracy concerning multiple relevant codes for a query, researchers adopted  Precision@k, MAP@k, and NDCG@k, where Precision@k equals to the percentage of relevant code in the top-k result list; MAP@k considers Precision@k with k ranging from 1 to k; NDCG@k is also a position sensitive metric similar to MAP@k but implemented in terms of the discounted cumulative gain. 

The previous metrics are based on code-query relevancy status (i.e., relevant or not). In contrast, we found that one study measures code search tool performance based on relevancy scores like MSE@k, the mean squared errors between the relevancy ratings by a tool, and the ratings by researchers or developers. Similarly, the correlation test between two ratings lists can also be used as a metric, including the KCC@k (Kendall's correlation coefficient \cite{abdi2007kendall}), SCC@k (Spearman correlation coefficient \cite{myers2004s}), and PCC@k (Pearson correlation coefficient \cite{benesty2009pearson}).

Some studies regard  code search as a classification task. Different from treat it as a ranking task, classification ignores the position of located code snippets for a given query and aims to retrieve as many relevant code snippets as possible \cite{chen2019capturing,ding2016kam1n0,lee2010instant}. For a query, a code search tool searches all relevant code from the codebase without size limitation. Its performance can then be measured by Precision (the proportion of code in a result list is truly relevant for a query) and Recall (the proportion of relevant code in the codebase that is correctly retrieved for a query). 

However, the same tool usually cannot usually achieve both high Precision and high Recall. To solve this issue, studies have also used summary metrics: F1-score, the harmonic mean of Precision and Recall \cite{lee2010instant,liu2019two,liu2018cross}; F2-score, a variant of F1-score that puts more weights on Precision \cite{khoo2013rendezvous}; AUROC, the area under the receiver operating characteristics curve \cite{linstead2009sourcerer}; and AUPR, the area under the precision-recall curve \cite{ding2016kam1n0}. Despite the accuracy measures, 13 studies also considered the tool efficiency in terms of the code search time \cite{white2016deep,sajnani2016sourcerercc}.

\begin{framed}
\filbreak
\noindent\textit{\textbf{Summary of answers to RQ3:}}
\begin{itemize}
\item\textit{Most reviewed study codebases were built with large-scale method-level source code written in Java, which were collected from public code repositories (e.g., GitHub and FDroid).}\vspace{3pt}
\item\textit{Most code search studies have tested their proposed tools with top-n frequently used text-based queries collected from Q\&A forums (e.g., Stack Overflow).}\vspace{3pt}
\item\textit{Performance of 55\% of code search tools were estimated using manual analysis.}\vspace{3pt}
\item\textit{Most of the reviewed studies assessed tool performance with ranking metrics (e.g., MRR and NDCG).}
\end{itemize}
\end{framed}


\section{Challenges and Opportunities}\label{challenge}

\subsection{Challenges}

\vspace{5pt}\noindent\textbf{Challenge 1: Diversity of the Codebase.} The characteristics of a codebase determine what a tool can find in the search space. However, most researchers have built their codebases in different ways that may affect the tool performance and usability: 

\textit{1) Small Scale.} The codebase scale in a practical environment (e.g., GitHub and FDroid) is usually large with millions of lines of code. However, many code search studies only tested their tools on a small scale codebase \cite{huang2020code,yang2017iecs,li2019reinforcement,wang2013mining}. In this case, findings may not generalize to large codebases \cite{li2019reinforcement}. Moreover, for a small-scale codebase, a code search tool may not work just because the codebase contains no code relevant to some search queries. 

\textit{2) Language Specific.} As illustrated in Fig. \ref{fig_codebase_language_scale}, most codebases only focused on code written in one type of programming language, especially Java. However, developers write and search code in various programming languages (e.g., Python and C\#). Although some studies claimed that their tools can be easily extended to other languages, they only tested their tools on a codebase for one specific language \cite{huang2020code,wan2019multi,sun2019slicing}.

\vspace{5pt}\noindent\textbf{Challenge 2: Limited Queries.} To test the tool effectiveness, studies carefully collected queries from Q\&A forums, development kits, and frequently used examples as listed in Table \ref{tab_query_source}. The studies tried their best to find appropriate queries to simulate developers' search behaviors in the real world. However, the selected queries are limited in four aspects:

\textit{1) Limited Quantity.} Fig. \ref{fig_query_type_scale}(b) shows that nearly 60\% of studies tested their proposed tools by using no more than 100 queries. Such query scale can only cover a limited number of real-world queries actually used by software developers \cite{huang2020code,hu2020unsupervised,sirres2018augmenting}. The main reason that hinders the query scale is the tool evaluation method. Table \ref{tab_evaluation_method} shows that 55\% of the code search tools were verified by manual evaluation only. Increasing the query scale would substantially increase the burden of user study participants during their labelling efforts.

\textit{2) Query Representativeness.} Commonly, code search studies selected the top-$n$ frequently used queries from Q\&A forums, or randomly sample a small set of queries from the real world \cite{huang2020code,hu2020unsupervised,sirres2018augmenting}. However, studies seldom investigated the representativeness of the selected queries. Thus, it would be uncertain and questionable if a code search tool can work for other types of queries. Moreover, the selected queries are usually too general and in reality developers often search for very domain-specific code \cite{wang2013mining}. Therefore, it is necessary to analyze the distribution and characteristics of queries to verify the tool generalizability. 

\textit{3) English Query Only.} For text-based queries, code search studies have nearly always only investigated the queries written in English. However, developers are scattered all around the world and use different languages, not just English \cite{sirres2018augmenting}. Therefore, it is beneficial and necessary to make an extension to non-English queries especially for text-based search tools. It is also necessary to investigate how developers with more limited English can use English-based code search tools.

\vspace{5pt}\noindent\textbf{Challenge 3: Model Construction Issues.} Learning models are popular and favored in recent years. This is because learning models (e.g., deep learning models) require no substantial manual efforts in incorporating domain knowledge into the traditional and heuristic models. However, existing learning models possess several threats to their model validity:

\textit{1) Validity of Parameters.} DL models involve many internal parameters, such as batch size, stop condition, learning rate, etc.  Code search studies have usually initialized these parameters with default settings and do not verify the effectiveness of this parameter choice \cite{huang2019enhance,huang2019qe,nie2016query}. They have also sometimes tuned the parameters without explaining the reasons or validity. Such unverified parameters may threaten the model generalizability \cite{liu2020replicability}.

\textit{2) Quality and Quantity of Training Data.} For text-based code search, DL models were trained with pairs of code and comments. The comment is a replacement for the query. Nevertheless, developers wrote queries in different styles and languages. Noisy comments also likely affect model performance. Thus, the quality of this training data may adversely threaten the trained model effectiveness \cite{shuai2020improving,yao2019coacor,wan2019multi}. Moreover, only a few codes contain comments so that a model trained with commented code may not work for other code \cite{liu2020simplifying}.

\vspace{5pt}\noindent\textbf{Challenge 4: Evaluation Issues.} We observed that tool evaluation is the most widely discussed future work issue in the reviewed code search studies. Based on their discussions and our findings, we attributed this evaluation issue to two aspects:

\textit{1) Relevancy Identification.} Table \ref{tab_evaluation_method} shows that only 28\% of code search tools were evaluated by automated identification with a carefully curated ground-truth. This is because a ground-truth is difficult to construct for most code search tasks. Although nine code search studies identified the query-code relevancy by designed measurements, it is uncertain if such measurements are actually reasonable. Due to the above difficulties, most code search studies chose to identify relevancy with human efforts. However, their evaluation with manual identification is subject to potential serious bias and human errors \cite{liu2020replicability,hu2020unsupervised,wan2019multi,liu2018supporting,gu2018deep}. 

\textit{2) Dataset Configuration.} For a tool based on a learning model, studies commonly split the codebase into three parts (i.e., training, validating, and testing data) with different ratios (e.g., 8:1:1) \cite{huang2019qe,chen2019capturing}. This is a common setting for tool verification and the split can avoid the overfitting issue \cite{gu2018deep}. However, this setting substantially reduces the scale of the codebase in tool testing. In a real-world code search scenario, the search space usually contains millions of repositories with limited number of training data. The testing data scale is also increasing continuously. Therefore, to better simulate practical search cases, it is suggested to split the codebase with less training data and more testing data.

\vspace{5pt}\noindent\textbf{Challenge 5: Limited Performance Measures.} Although researchers used various performance metrics to measure the tool performance, as listed in Table \ref{tab_performance}, the adopted metrics are not enough to meet the requirements for code search evaluation. We observed that current code search tools lack the following considerations:

\textit{1) Tool Efficiency and Scalability.} Searching for relevant code from a large-scale codebase is one feature of the code search task. Therefore, developers expect that the used tool performs code search as fast as possible. However, most code search studies did not estimate the tool performance in terms of the tool search time, as illustrated in Table \ref{tab_performance}. Especially for tools based on learning models, the search time is commonly not acceptable due to the high model complexity \cite{liu2020simplifying,fu2017easy}. Moreover, as the codebase is frequently updated by developers, it is also necessary to estimate the scalability and reproducibility of the proposed tool on codebases of different scales.
    
\textit{2) Other Important Metrics.} To estimate the performance of code search tools, the accuracy (e.g., MRR and NDCG) and efficiency (e.g., search time) metrics are not enough. Some code search studies (e.g., text-based search, API-based search, and exampled search) just return a list of relevant code to a search query, but ignore the diversity of the list of returned code without excluding repeated results \cite{martie2015sameness}. Some returned code snippets are not concise with many lines irrelevant to the actual query intent. This kind of code can not be reused easily by developers in their real-world software development scenarios \cite{zhang2019enriching}. Therefore, it is important to consider the readability and reusability of the searched code \cite{moreno2015how}.

\vspace{5pt}\noindent\textbf{Challenge 6: Replication Issues.} Tool reproducibility is of high merit in code search research, as other researchers or practitioners require less effort in replicating / extending / comparing to the study. We observed that existing code search tools suffer from the following replication issues:

\textit{1) Replication Package.} Sharing source code and dataset can greatly help to support the tool replicability and mitigate researchers' replication efforts. However, Fig. \ref{fig_replicability} shows that only 18\% of code search tools provide accessible replication packages \cite{liu2020replicability,sirres2018augmenting}. It is recommended that future studies share their contributions publicly \cite{liu2020replicability}. Some researchers may share their code upon email request, but it is highly recommended to provide the accessible replication package links in the papers and to maintain these repositories.    

\textit{2) Online Search Engine.} Table \ref{tab_model} shows that ten code search tools depend on an online search engine (e.g., GitHub search, Google Code search, and Sourcerer). However, the online search engines could be improved through time, so that new studies cannot replicate the experimental results reported by early studies. To facilitate the tool replication, further studies are encouraged to provide the access date of their used online search engines. Besides, if there exists source code or paper for an online engine (e.g., Sourcerer), it is necessary to check their differences. In this way, when researchers found the performance of their re-implemented code search tool is different from the performance reported in a paper, they could understand whether the difference comes from the improvement of the online search engine or their re-implementation errors.

\subsection{Opportunities}

\vspace{5pt}\noindent\textbf{Opportunity 1: Better Benchmarks.} One urgent task for code search research is to build a standard benchmark that different tools can be evaluated with. Such a benchmark requires that: the codebase is industrial-scale with millions of lines of code and multiple programming languages \cite{huang2020code,wan2019multi}; the tested search queries involve various diverse types of real-world examples covering not just top-$n$ frequently used general queries; developing a standard automated tool evaluation method to exclude manual efforts and subjective bias; and excluding repeated code before assessing the tool performance. 

\vspace{5pt}\noindent\textbf{Opportunity 2: DL-Based Model with Big Data.} It is promising to build DL-based code search models trained with big data to learn the correlations between queries and code. There are some opportunities to improve the DL-based models such as: standardizing code written by different developers with various programming styles and experiences to mitigate training difficulties; improving the quality of the training data; developing better code representation methods to capture the programming semantics \cite{zhang2019novel,shuai2020improving}; leveraging better loss functions to optimize the DL-based model, e.g., using the ranking loss function \cite{shuai2020improving,ye2020leveraging,liu2018recommending,liu2018recommender}, and increasing size of the training and testing data sets.

\vspace{5pt}\noindent\textbf{Opportunity 3: Fusion of Different Types of Models.} Researchers have developed various code search tools. Although learning models have shown promising advantages over traditional and heuristic models, their disadvantages are still obvious, such as slow code search time,  low scalability and need for periodic retraining. However such efficiency issues can be complemented by the other two types of models. Therefore, it is suggested to explore fusing the advantages of different types of code search models into tools \cite{liu2020simplifying}.

\vspace{5pt}\noindent\textbf{Opportunity 4: Multi-Language Tool.} It is difficult to apply an existing tool designed for a specific programming language to searching code written in other programming languages. This is because tools often depend on language-specific features and parsing processes. Moreover, a learning model is likely to cost a long time to retrain data for a new programming language. Therefore, it is recommended to develop a multi-language code search tool. For example, leveraging the multi-task learning techniques \cite{zhang2018overview,zhang2017survey} to capture the semantics of multi-language in code search, and investigating whether a model trained on one language can be transferred to other language \cite{pan2009survey,torrey2010transfer}. In this way, code search practitioners do not have to train and deploy code search tools on each programming language one by one.

\vspace{5pt}\noindent\textbf{Opportunity 5: New Code Search Tasks.} 
UI code search \cite{xie2019user,behrang2018guifetch,reiss2018seeking} and programming video search \cite{bao2020psc2code} are two emerging code search tasks. Different from the traditional popular code search tasks using text-based code search, they extended the capability of existing code search engines to use UI sketches as queries and videos as sources for useful code snippets.  Specifically, the UI code search can provide more relevant help for UI developers, while programming video search provides more detailed tutorials for novice developers. Therefore, it is recommended that researchers pay more attentions for new tools for these new code search tasks.



\section{threats to Validity}\label{threat}
\subsection{Publication Bias}
Publication bias indicates the issue of publishing more positive results over negative results \cite{keele2007guidelines}. This is because positive results, e.g., a code search tool with statistically significant advantages over baselines, have a much higher chance of getting published. Meanwhile, negative results, e.g., the suspected flawed studies, are likely rejected for publication. Thus, to ensure the publication chance, some studies may report biased, incomplete and incorrect conclusions due to their low quality of experimental design (e.g., using limited or selected testing data). Therefore, the claims in this review supported or rejected by our selected major studies could be biased if the original literature suffers from such publication bias. 

\subsection{Search Terms}
It is always challenging to find all relevant primary studies in any systematic review \cite{keele2007guidelines,petersen2015guidelines}. To address this issue, we presented a detailed search strategy. The search string was constructed with terms identified by checking titles, abstracts, and keywords from many relevant papers that were already known to the authors. The adopted search string was piloted and the identified studies confirmed the applicability of the search string. These procedures provided high confidence that the majority of the key studies were included in the review.

\subsection{Study Selection Bias}
The study selection process was carried out in two phases. The first phase excluded studies based on the title and abstract by two independent researchers. A pilot study of the selection process was conducted to place a foundation for a better understanding of the inclusion/exclusion criteria. Potential disagreements were resolved during the pilot study and inclusion/exclusion criteria were refined. Inter-rater reliability was evaluated to mitigate the threat that emerged from the researchers' personal subjective judgment. The agreement between the two researchers was "substantial" for selecting relevant papers from the full set of papers. The selection process was repeated until a full agreement was achieved. When the researchers could not decide on a particular study, a third researcher was consulted. The second phase was based on the full text. Due to this well-established study selection process, it is unlikely that any relevant studies were missed.

\subsection{Data Extraction}
For data extraction, the studies were divided between two researchers; each researcher extracted the data from the relevant studies and the extracted data were rechecked by the other researcher. Issues in data extraction were discussed after the pilot data extraction and the researchers were able to complete the data extraction process following the refinement of the criteria. Extracted data were then inspected by automated scripts to check the correctness of the extracted values across the paper content, improving the validity of our analysis.

\subsection{Data Analysis}

We may have mis-compared some studies, mis-understood some of the techniques or evaluation methods reported, missed replication package links or failed to find repositories searching with paper title, or may have mis-understood reported dataset and evaluation metric details. 
From the extracted data we took due care in each of these areas to properly analyse, represent, classify and summarise the reviewed studies in this paper. However, there may be errors in our classification and analysis that impacts the overall findings of this review.

\section{conclusion}\label{conclude}
In recent years, researchers have proposed many tools to support the very common and important code search task to help boost developers' software development productivity and quality of code produced. To investigate the current state of research on code search, we performed a comprehensive review of 81 code search studies found from searching three electronic databases, ACM Digital Library, IEEEXplore, and ISI Web of Science. After extracting data from and analysing these selected studies, we found that:

\begin{itemize}
    \item The popularity of code search research is substantially increasing with a peak in 2019, which accounts for more than 50\% of the studies published from 2002-2020; 60\% of studies were published in conferences; 83\% of the studies proposed new code search tools. \vspace{3pt}
    
    \item 74\% of the reviewed tools focused on the general code search task (inputted with text-based queries, API-based queries, and input/output example) to improve the existing code search engines (e.g., GitHub Search); 21\% of the tools aimed to search similar source/binary code from codebase; the remaining 5\% of tools intended to search UI code or code in programming videos; in the recent two years, researchers frequently leverage DL techniques to tackle the challenges in early tools; but only 18\% of code search tools shared accessible replication package links in their papers or provided source code in GitHub. \vspace{3pt}
    
    \item To verify tool validity, method-level Java code collected from open source community (e.g., GitHub and FDroid) is the first choice to build a large-scale codebase; Representative search queries were commonly extracted from Q\&A forums, development kit, and frequently used examples; most studies manually identified the relevancy of query and searched code and assessed the tool performance by using the ranking metrics (e.g., MRR and NDCG). 
\end{itemize}

After analyzing the shortcomings of existing code search tools, we recommend further studies to build a better benchmark with large-scale code written in multiple programming languages, including various queries covering not only top frequently used examples but also domain-specific cases, and a standard automated tool evaluation method. DL-based tools require further improvements, such as better code representation, higher quality of training data, and speed improvements. It is recommended to fuse them with other models such as traditional IR-based and heuristic models. It is difficult to deploy an existing tool to search code written in multiple programming languages. Therefore, a multi-language tool is needed in the future. Additionally, new code search tasks such as searching UI code or code in programming videos are also worthy of more attention.


\bibliographystyle{ACM-Reference-Format}
\bibliography{reference}

\end{document}